\documentstyle[preprint,tighten,aps,floats,epsf,psfig,amssymb]{revtex}

\def\smfrac#1#2{{\textstyle\frac{#1}{#2}}}   

\begin{document}
\draft
\title{
Critical exponents and equation of state of
the three-dimensional Heisenberg universality class
}

\author{Massimo Campostrini,${}^{1,}$\cite{MC-email}
        Martin Hasenbusch,${}^{2,}$\cite{MH-email}
        Andrea Pelissetto,${}^{3,}$\cite{AP-email}\\
        Paolo Rossi,${}^{1,}$\cite{PR-email} and
        Ettore Vicari${}^{1,}$\cite{EV-email}}
\address{${}^1$ Dipartimento di Fisica dell'Universit\`a di Pisa 
and I.N.F.N., I-56126 Pisa, Italy}
\address{${}^2$ NIC/DESY Zeuthen, 
Platanenallee 6, D-15738 Zeuthen, Germany}
\address{${}^3$ Dipartimento di Fisica dell'Universit\`a di Roma I
and I.N.F.N., I-00185 Roma, Italy
}


\preprint{DESY 01-154, IFUP-TH 2001/33}

\maketitle

\begin{abstract}
We improve the theoretical estimates of the critical exponents for the
three-dimensional Heisenberg universality class.  
We find $\gamma=1.3960(9)$, $\nu=0.7112(5)$, $\eta=0.0375(5)$,
$\alpha=-0.1336(15)$, $\beta=0.3689(3)$, and $\delta=4.783(3)$.  
We consider an improved lattice $\phi^4$ Hamiltonian
with suppressed leading scaling corrections.
Our results are obtained by combining Monte Carlo
simulations based on finite-size scaling methods and high-temperature
expansions.   The critical
exponents are computed from high-temperature expansions specialized to
the $\phi^4$  improved model.  
By the same technique we determine the
coefficients of the small-magnetization expansion of the equation of
state.  This expansion is extended analytically by means of
approximate parametric representations, obtaining the equation of
state in the whole critical region.  We also determine a
number of universal amplitude ratios.
\end{abstract}



\section{Introduction and summary}
\label{introduction}
According to the universality hypothesis, some features of continuous phase
transitions---for instance, critical exponents and scaling functions---do 
not depend on the microscopic details of the systems, but
only on few global properties, 
such as the space dimensionality, the range of interaction, and the
symmetry of the order parameter. These features define  a universality 
class. In this paper,
we consider the three-dimensional Heisenberg universality class, which is
characterized by a three-component order parameter, ${\rm O}(3)$
symmetry, and short-range interactions.

The Heisenberg universality class describes \cite{foot1} the critical 
behavior of isotropic
magnets, for instance the Curie transition in isotropic ferromagnets 
such as Ni  and EuO, and of antiferromagnets such as 
RbMnF$_3$ at the N\'eel transition point. 
In Table \ref{table_exponents_exp_1} we report 
some recent experimental results. It is not 
a complete review of published results, but it is
useful to get an overview of the experimental state of the art.  
In the table we have also included results for the well-studied 
doped manganese perovskites La$_{1-x}$A$_{x}$MnO$_3$, although the 
nature of the ferromagnetic transition in these compounds is still unclear
\cite{foot2}. The Heisenberg universality class also describes isotropic 
magnets with quenched disorder. Indeed, since $\alpha < 0$, the Harris
criterion states that disorder is an irrelevant 
perturbation. The only effect is to introduce 
a correction-to-scaling term $|t|^{\Delta_{\rm dis}}$ with 
$\Delta_{\rm dis} = - \alpha$. The experimental 
results confirm the theoretical analysis \cite{foot3}, as it can be seen 
from Table \ref{table_exponents_exp_2} (older experimental results with 
a critical discussion are reported in Ref. \cite{Kaul-85}). The 
prediction for $\Delta_{\rm dis}$ has been checked in 
perturbative field theory \cite{PV-random} and experimentally
\cite{Kaul-88,RK-95,RK-95b}.

\begin{table}[tp]
\caption{\label{table_exponents_exp_1}
Recent experimental estimates of the critical exponents for Heisenberg systems.}
\begin{tabular}{lllll}
&
\multicolumn{1}{c}{Material}&
\multicolumn{1}{c}{$\gamma$}&
\multicolumn{1}{c}{$\beta$}&
\multicolumn{1}{c}{$\delta$}\\
\tableline \hline
Ref. \cite{SBEW-80} (1980)        & Ni     &           & 0.354(14) &       \\
Ref. \cite{Kobeissi-81} (1981)    & Fe     &           & 0.367(5)  &       \\
Ref. \cite{Seeger-etal-95} (1995) & Ni     & 1.345(10) & 0.395(10) & 4.35(6) \\
Ref. \cite{RKS-95} (1995)     & Gd$_2$BrC  & 1.392(8)  & 0.365(5)  &4.80(25) \\
Ref. \cite{RKS-95} (1995)     & Gd$_2$IC   & 1.370(8)  & 0.375(8)  &4.68(25) \\
Ref. \cite{Zhao-etal-99} (1999) & Tl$_2$Mn$_2$O$_7$ 
                                           & 1.31(5)   & 0.44(6)   &4.65(15) \\
Ref. \cite{Barsov-etal-00} (2000) & La$_{0.82}$Ca$_{0.18}$MnO$_3$ 
                                           &           & 0.383(9)  &         \\
Ref. \cite{Zhao-etal-00} (2000) &  La$_{0.95}$Ca$_{0.05}$MnO$_3$  
                                           & 1.39(5)   & 0.36(7)   &4.75(15) \\
Ref. \cite{ArPa-00} (2000)    & Gd(0001)   &           & 0.376(15) &         \\
Ref. \cite{MRBSGV-00} (2000)  & Gd$_2$CuO$_4$ & 1.32(2)& 0.34(1)   &         \\
Ref. \cite{Buhrer-etal-00} (2000) & C$_{80}$Pd$_{20}$ (liq) & 1.42(5) & &    \\
Ref. \cite{Buhrer-etal-00} (2000) & C$_{80}$Pd$_{20}$ (sol) & 1.40(8) & &    \\
Ref. \cite{Brueckel-etal-01} (2001) & GdS  &           & 0.38(2)   &         \\
Ref. \cite{Yang-etal-01} (2001) & CrO$_2$  & 1.43(1)   & 0.371(5)  &         \\
Ref. \cite{HKH-01} (2001)  &  La$_{0.8}$Ca$_{0.2}$MnO$_3$ 
            & 1.45    & 0.36     & \\
\end{tabular}
\end{table}

\begin{table}[tbp]
\caption{\label{table_exponents_exp_2}
Recent experimental estimates of the critical exponents for Heisenberg
systems with quenched disorder.}
\begin{tabular}{lllll}
&
\multicolumn{1}{c}{Material}&
\multicolumn{1}{c}{$\gamma$}&
\multicolumn{1}{c}{$\beta$}&
\multicolumn{1}{c}{$\delta$}\\
\tableline \hline
Ref. \cite{KR-94} (1994) & Fe$_{10}$Ni$_{70}$Bi$_{19}$Si 
                  &    1.387(12)   & 0.378(15)   & 4.50(5)   \\
Ref. \cite{KR-94} (1994) & Fe$_{13}$Ni$_{67}$Bi$_{19}$Si 
                  &    1.386(12)   & 0.367(15)   & 4.50(5)   \\
Ref. \cite{KR-94} (1994) & Fe$_{16}$Ni$_{64}$Bi$_{19}$Si 
                  &    1.386(14)   & 0.360(15)   & 4.86(4)   \\
Refs. \cite{RK-95,RK-95b} (1995) & Fe$_{20}$Ni$_{60}$P$_{14}$B$_6$
                  &    1.386(10)   & 0.367(10)   & 4.77(5)   \\
Refs. \cite{RK-95,RK-95b} (1995) & Fe$_{40}$Ni$_{40}$P$_{14}$B$_6$
                  &    1.385(10)   & 0.364(5)    & 4.79(5)   \\
Ref. \cite{BK-97} (1997) & Fe$_{91}$Zr$_9$ 
                  &    1.383(4)    & 0.366(4)    & 4.75(5)   \\ 
Ref. \cite{BK-97} (1997) & Fe$_{89}$CoZr$_{10}$ 
                  &    1.385(5)    & 0.368(6)    & 4.80(4)   \\
Ref. \cite{BK-97} (1997) & Fe$_{88}$Co$_2$Zr$_{10}$ 
                  &    1.389(6)    & 0.363(5)    & 4.81(5)   \\
Ref. \cite{BK-97} (1997) & Fe$_{84}$Co$_6$Zr$_{10}$ 
                  &    1.386(6)    & 0.370(5)    & 4.84(5)   \\
Ref. \cite{SHAM-99} (1999) & Fe$_{1.85}$Mn$_{1.15}$Si 
                  &    1.543(20)   & 0.408(60)   & 4.74(7)   \\
Ref. \cite{SHAM-99} (1999) & Fe$_{1.50}$Mn$_{1.50}$Si 
                  & 1.274(60)      & 0.383(10)   & 4.45(19)  \\
Ref. \cite{Perumal-etal-00} (2000) & Fe$_{86}$Mn$_4$Zr$_{10}$
                  & 1.381          & 0.361       & \\
Ref. \cite{Perumal-etal-00} (2000) & Fe$_{82}$Mn$_8$Zr$_{10}$
                  & 1.367          & 0.363       & \\
Ref. \cite{PSKYRD-01} (2001) & Fe$_{84}$Mn$_6$Zr$_{10}$
                  & 1.37(3)        & 0.359       & 4.81(4)    \\
Ref. \cite{PSKYRD-01} (2001) & Fe$_{74}$Mn$_{16}$Zr$_{10}$
                  & 1.39(5)        & 0.361       & 4.86(3)    \\
\end{tabular}
\end{table}

Beside the exponents $\gamma$, $\beta$, and $\delta$ there are
also a few estimates of the specific-heat exponent $\alpha$, in most 
of the cases obtained from resistivity measurements: 
$\alpha \approx -0.10$ in Fe and Ni \cite{KHM-81};
$\alpha = -0.12(2)$ in EuO \cite{SWSLPK-92}; 
$\alpha = -0.11(1)$ in Fe$_x$Ni$_{80-x}$B$_{19}$Si \cite{KR-94};
$\alpha = -0.11(1)$ in RbMnF$_3$ \cite{MMFB-96}.	

Aim of this paper is to substantially improve the precision of the
theoretical estimates of the critical exponents. For this purpose, we 
consider an improved lattice Hamiltonian that is characterized by
the fact that the leading correction to scaling is (approximately) absent in the
expansion of any observable near the critical point. Moreover, we
combine Monte Carlo (MC) simulations and analyses of 
high-temperature (HT) series. We exploit the
effectiveness of MC simulations and
finite-size scaling (FSS) techniques
to determine the critical temperature and the parameters of the improved
Hamiltonians 
\cite{BFMM-98,BFMMPR-99,HPV-99,Hasenbusch-99,HT-99,Habil,%
CHPRV-01,Hasenbusch-00},
and the effectiveness of HT methods to determine the
critical exponents for improved models, especially when a precise
estimate of the critical point is available.  
This approach has already been applied to the
three-dimensional Ising \cite{CPRV-99} and $XY$  \cite{CPRV-00,CHPRV-01} 
universality classes, achieving a substantial improvement
of the estimates of the universal quantities that describe the
critical behavior, such as the critical exponents and 
the scaling equation of state. 

We consider a simple cubic lattice and 
the nearest-neighbor $\phi^4$ lattice Hamiltonian 
\begin{equation}
{\cal H}_{\phi^4} =
 - \beta\sum_{\left<xy\right>} {\vec\phi}_x\cdot{\vec\phi}_y +\, 
   \sum_x \left[ {\vec\phi}_x^{\,2} + \lambda ({\vec\phi}_x^{\,2} - 1)^2\right],
\label{phi4Hamiltonian}
\end{equation}
where $\vec{\phi}_x$ is a three-component field.
As shown in Ref.~\cite{Hasenbusch-00}, 
the Hamiltonian (\ref{phi4Hamiltonian}) is improved  for 
$\lambda = \lambda^* \approx 4.4(7)$.
Here, we extend the simulations of Ref.~\cite{Hasenbusch-00}, obtaining 
a more accurate estimate of $\lambda^*$,
$\lambda^*=4.6(4)$,  and  precise estimates of the critical $\beta_c$ 
for several values of $\lambda$.
The analysis of the MC FSS results obtained for the 
improved $\phi^4$ lattice Hamiltonian already
provides precise estimates of the critical exponents.  
As shown in Refs.~\cite{CPRV-99,CPRV-00,CHPRV-01}, an additional
increase in precision can be obtained by combining improved 
Hamiltonians and HT methods.  For this purpose,
by using the linked-cluster expansion technique, we
computed HT expansions of several quantities and
analyzed them using the MC results for $\lambda^*$ and
$\beta_c$. The final results significantly improve 
those obtained from the MC simulation. 
Moreover, they substantially improve 
those  obtained using longer (21 orders) series for the standard 
Heisenberg model \cite{BC-97}.

In Table \ref{table_exponents} we report our results for the critical
exponents. We give the estimates obtained from the analysis of the MC 
data alone and those obtained by combining MC and HT
techniques---they are denoted by MC+IHT, where the ``I" refers 
to the fact that we are considering an improved model.
The exponent $\alpha$ can be derived using the hyperscaling relation
$\alpha = 2 - 3\nu$, obtaining $\alpha=-0.1336(15)$.
We would like to stress  that the good
agreement between the MC and HT estimates is not trivial,
since the critical exponents are determined from 
different quantities and limits. Indeed,
the MC estimates are obtained from the 
analysis of the finite-size behavior for the size 
$L\rightarrow\infty$ at the critical point $\beta = \beta_c$, 
while the HT results are derived from the singular behavior 
of infinite-volume quantities as $\beta\to \beta_c$. 

In Table \ref{table_exponents} we 
compare our results with the most precise theoretical
estimates that have been obtained in recent years.  
A more complete list of results can be found in Ref.~\cite{PV-review}.
The results we quote have been obtained by Monte Carlo simulations (MC),
from the analysis of the HT series
for the standard Heisenberg model (HT),  or by  field-theory methods (FT).  
The MC results were obtained by applying FSS techniques to different
Hamiltonians. Refs. \cite{BFMM-96,HJ-93,PFL-91} studied 
the standard $O(3)$-vector Heisenberg model,
Ref.~\cite{Hasenbusch-00} 
the improved $\phi^4$ model, and Ref.~\cite{CBL-00} 
an isotropic ferromagnet with double-exchange interactions
\cite{foot-doubleexchange}.
The HT results of Ref.\ \cite{BC-97} 
were obtained analyzing the 21st-order HT expansions for the standard
O(3)-vector model on the simple cubic (sc) and on the 
body-centered cubic (bcc) lattice.
The FT results of Refs.~\cite{JK-01,GZ-98,MN-91,LZ-77,YG-98,Kleinert-00} 
were derived 
by analyzing perturbative expansions in different frameworks:
fixed-dimension expansion (6th- and 7th-order series, see 
Refs. \cite{BNGM-77,MN-91}),
$\epsilon$-expansion (to $O(\epsilon^5)$, see Refs. \cite{CGLT-83,KNSCL-93}),
and $(d-2)$-expansion (to $O[(d-2)^4]$, see Refs. \cite{HB-78,Hikami-83,BW-86}). 
We quote two errors for the results of Ref.~\cite{MN-91}: 
the first one (in parentheses) 
is the resummation error, and the second one (in brackets) 
takes into account the uncertainty of the fixed-point value $g^*$ of the
coupling, which was estimated to be approximately 1\% in Ref.~\cite{MN-91}. 
To estimate the second error we use the results of
Ref.~\cite{GZ-98} where the dependence of the exponents on $g^*$ is given.
The results of Ref.~\cite{NR-84} were obtained by using the so-called 
scaling-field method (SFM). 
Refs.~\cite{BSW-01,GW-01,BTW-96,BTW-00} present results
obtained by approximately solving continuous renormalization-group
(CRG) equations for the average action, which is approximated
to lowest and first order of the derivative expansion.
We also mention the HT results of Ref. \cite{BC-99}:
they performed a direct determination of the exponent $\alpha$ obtaining 
$\alpha = -0.11(2)$, $-0.13(2)$ on the sc and bcc
lattice. Ref. \cite{MOF-01} computes the critical exponents for a Heisenberg
fluid by a canonical-ensemble simulation. Depending on the analysis method, 
they find $1/\nu = 1.40(1)$, $1.31(1)$, $\beta/\nu= 0.54(2)$, $0.52(1)$, 
and  $\gamma/\nu =1.90(3)$, $1.87(3)$.
Overall, all estimates are in substantial agreement with our MC+IHT
results. We only note the apparent discrepancies with  
the MC estimates of $\eta$ of Refs.~\cite{BFMM-96,HJ-93},
and with the FT results of Ref. \cite{JK-01}. 
However, the reliability of the
error bars reported in Ref.~\cite{JK-01} is unclear: indeed, 
Ref. \cite{GZ-98} analyzes the 
same perturbative series and reports much more cautious error estimates.

\begin{table}[tp]
\caption{\label{table_exponents}
Estimates of the critical exponents. See text for the explanation of
the symbols in the second column. We indicate with an asterisk (${}^*$)
the estimates that have been obtained using the relations
$\gamma = (2-\eta)\nu$, $2\beta=\nu(1+\eta)$, $\delta (1+\eta) = 5-\eta$.
}

\begin{tabular}{rclllll}
\multicolumn{1}{c}{Ref.}&
\multicolumn{1}{c}{Method}&
\multicolumn{1}{c}{$\gamma$}&
\multicolumn{1}{c}{$\nu$}&
\multicolumn{1}{c}{$\eta$}&
\multicolumn{1}{c}{$\beta$}&
\multicolumn{1}{c}{$\delta$}\\
\tableline \hline
\multicolumn{1}{c}{this work} & 
MC+IHT &1.3960(9)      &0.7112(5)   & 0.0375(5)  & 0.3689(3)$^*$ &4.783(3)$^*$\\
\multicolumn{1}{c}{this work} &  
MC    &1.3957(22)$^*$ &0.7113(11)  & 0.0378(6)  & 0.3691(6)$^*$ &4.781(3)$^*$ \\
\cite{Hasenbusch-00} (2000)   &  
MC    &1.393(4)$^*$   &0.710(2)    & 0.0380(10) & 0.3685(11)$^*$ & 4.780(6)$^*$\\
\cite{CBL-00}  (2000)         &  
MC    &1.3909(30)  &0.6949(38)  & & 0.3535(30) & \\
\cite{BFMM-96} (1996)         &  
MC    &1.396(3)$^*$   &0.7128(14)  & 0.0413(16) & 0.3711(9)$^*$ & 4.762(9)$^*$\\
\cite{HJ-93}   (1993)         &  
MC    &1.389(14)$^*$  &0.704(6)    & 0.027(2)    & 0.362(3)$^*$  & 4.842(11)$^*$\\
\cite{PFL-91}  (1991)         &  
MC    &1.390(23)$^*$  &0.706(9)    & 0.031(7)    & 0.364(5)$^*$  & 4.82(4)$^*$\\
\cite{BC-97} (1997)         & 
HT sc  & 1.406(3)      & 0.716(2)   & 0.036(7)$^*$& 0.3710(13)$^*$ &4.79(4)$^*$ \\
\cite{BC-97} (1997)         & 
HT bcc & 1.402(3)      & 0.714(2)  &  0.036(7)$^*$& 0.3700(13)$^*$ & 4.79(4)$^*$ \\
\cite{AHJ-93}  (1993)         & HT     & 1.40(1)       & 0.712(10) &  0.03(3)$^*$&  0.368(6)$^*$ \\
\cite{JK-01}   (2001) &
FT $d=3$ exp & 1.3882(10)& 0.7062(7)  & 0.0350(8) & 0.3655(5)$^*$  & 4.797(5)$^*$ \\
\cite{GZ-98}   (1998) & 
FT $d=3$ exp & 1.3895(50)& 0.7073(35) &0.0355(25) & 0.3662(25) & 4.794(14) \\
\cite{MN-91}   (1991) & 
FT $d=3$ exp & 1.3926(13)[39] & 0.7096(8)[22] &0.0374(4) & & \\
\cite{LZ-77}   (1977) & 
FT $d=3$ exp & 1.386(4)& 0.705(3) &0.033(4) & 0.3645(25) & 4.808(22) \\
\cite{GZ-98}   (1998) & 
FT $\epsilon$-exp & 1.382(9) & 0.7045(55) & 0.0375(45) & 0.3655(35) &4.783(25) 
\\
\cite{YG-98}   (1998) &
FT $\epsilon$-exp & 1.39$^*$ & 0.708 & 0.037 & 0.367$^*$ & 4.786$^*$ \\
\cite{Kleinert-00} (2000) & FT $(d-2)$-exp & & 0.695(10) & & \\
\cite{NR-84}  (1984) & 
SFM  & 1.40(3)   &  0.715(20)   & 0.044(7) & 0.373(11) & 4.75(4)$^*$\\
\cite{BSW-01} (2001) & CRG && 0.74 & 0.038 & 0.37 & 4.78 \\
\cite{GW-01} (2001)  & CRG & 1.374 & 0.704  & 0.049  &  0.369 & 4.720 \\
\cite{BTW-96} (1996)  & CRG & 1.465 & 0.747  & 0.038  &  0.388 & 4.78 \\
\end{tabular}
\end{table}

We also present a detailed study of the equation of state.  We first
consider its expansion in terms of the magnetization in the
HT phase.  The coefficients of this expansion are
directly related to the zero-momentum $n$-point renormalized
couplings, which are determined by analyzing their HT expansion.
These results are used to construct parametric representations of the
critical equation of state which are valid in the whole critical
region, satisfy the correct analytic properties (Griffiths'
analyticity), and take into account the Goldstone singularities at the
coexistence curve.  From our approximate representations of the
equation of state we derive estimates of several universal amplitude
ratios.  Moreover, we present several results and different forms of the 
equation of state that can be compared directly with experiments. 
In particular, we can compare with the experimental 
results of Refs.~\cite{KR-94,BK-97,Zhao-etal-99}, finding good 
agreement.

The paper is organized as follows.  In Sec.\ \ref{MonteCarlo} we
present our MC results.  
In Sec.\ \ref{HTanalysis} we present our results for the critical
exponents obtained from the analysis of the HT series 
for the improved Hamiltonian (\ref{phi4Hamiltonian}).  The equation
of state is discussed in Sec.\ \ref{CES}.  
We determine the small-magnetization expansion coefficients in
Sec.\ \ref{IHTrj}, give an approximate parametric representation 
of the equation of state in Secs. \ref{preq} and \ref{appeq},
compute several amplitude ratios in Sec. \ref{unratio}, and 
compare the theoretical results with experimental data in
Sec.\ \ref{CES.E}. Details are reported in the Appendices. 
In App. \ref{MonteCarlo-appendix} we present the analysis of the 
MC results and in App. \ref{seriesanalysis} the analysis of the 
HT series. The expressions of several amplitude ratios in terms
of the parametric representations are reported in App. \ref{univparrep}.

\section{Monte Carlo Simulations}
\label{MonteCarlo}

The present MC simulations extend those of Ref. \cite{Hasenbusch-00}.
Here, we have considerably enlarged the statistics and added
larger lattice sizes. Moreover, we have considered an additional quantity
in order to improve the control over systematic errors. This way, we can
increase the accuracy of $\lambda^*$ and 
give precise estimates of the critical $\beta_c$ 
for three values of $\lambda$ in a neighborhood of $\lambda^*$.
For a detailed discussion of our methods, see Ref. \cite{CHPRV-01}. 
Details are reported in App. \ref{MonteCarlo-appendix}.

We simulated
the $O(3)$-symmetric $\phi^4$ model (\ref{phi4Hamiltonian}) 
at $\lambda=4.0$, $4.5$, and $5.0$ on a simple cubic lattice
with linear extension $L$ in all directions. 
We measured the Binder parameter $U_4$, its sixth-order generalization
$U_6$, the second-moment correlation length $\xi_{\rm 2nd}$, and 
the ratio $Z_a/Z_p$, where
$Z_a$ is the partition function with anti-periodic boundary
conditions in one of the three directions and $Z_p$ the corresponding one
with periodic boundary conditions in all directions. 
The number of iterations for each lattice size and value of
$\lambda$ was approximately $10^7$ for $L=6,7,8,9,10,11,12,14,16,18,20,22$,
approximately $10^6$ for $L=24,28,32,36,40,48$, and
1-4$\times 10^5$ for $L=56,64,80,96$.
With respect to Ref. \cite{Hasenbusch-00}, 
we have added new lattice sizes for all three
values of $\lambda$ and considerably increased 
the statistics.
In total, the whole study took  about four years 
on a single 450 MHz Pentium III CPU.

In the first step of the analysis, we compute 
$\beta_c$ and the fixed-point value of 
the dimensionless ratios $R^*$ for $\lambda=4.5$, using the standard 
cumulant crossing method of Binder.
In particular, we fit our data with the ansatz
\begin{equation}
\label{bindercross}
R^* = R(L,\beta_c),
\end{equation}
where $R^*$ and $\beta_c$ are free parameters. 
Our results are reported in Table \ref{betacR}.
Note that the four results for $\beta_c$ are consistent within error bars.
The statistical error of $\beta_c$ obtained from $Z_a/Z_p$ and 
$\xi_{\rm 2nd}/L$ is
considerably smaller than that from $U_4$ and $U_6$. 
As our final estimate we take $\beta_c=0.6862385(20)$, which is consistent with 
all four results.

\begin{table}
\caption{\label{betacR} Final results for $\beta_c$ and $R^*$ 
from fits with ansatz (\ref{bindercross}). In parentheses 
we give the statistical error and in brackets the error 
due to the corrections to scaling.
}
\begin{tabular}{lllll}
   $R$ &  \multicolumn{1}{c}{$Z_a/Z_p$}    & 
          \multicolumn{1}{c}{$\xi_{\rm 2nd}/L$}  &
          \multicolumn{1}{c}{$U_4$}  & 
          \multicolumn{1}{c}{$U_6$} \\
\hline
\hline
 $R^*$    & 0.1944(1)[4]  & 0.5644(1)[2]& 1.1394(1)[2] & 1.4202(2)[10] \\
 $\beta_c$& 0.6862390(10)[12]& 0.6862386(11)[6]&0.6862365(17)[12]&0.6862369(17)[19]\\
\end{tabular}
\end{table}

In addition, we determine $\beta_c$ for $\lambda=4.0$ and $\lambda=5.0$. 
For this purpose, we use the ansatz (\ref{bindercross}), fixing  
$L=96$ and taking the values of $R^*$ from Table \ref{betacR}. Our results
are summarized in Table \ref{betacoff}. For both values of $\lambda$, 
the results obtained from the four different choices of $R^*$ are 
consistent within error bars.  As our final result we take that obtained
from $Z_a/Z_p$, since it has the smallest statistical error.

\begin{table}
\caption{\label{betacoff} Results for $\beta_c$ at $\lambda=4.0$ and $5.0$
using only $L=96$ and the ansatz $R(\beta_c)=R^*$, where $R^*$ is 
taken from Table \ref{betacR}. In parentheses we give the 
statistical error
and in brackets the error due to the uncertainty on $R^*$. 
}
\begin{tabular}{lllll}
         \multicolumn{1}{c}{$\lambda$}    & 
         \multicolumn{1}{c}{$Z_a/Z_p$}    & 
          \multicolumn{1}{c}{$\xi_{\rm 2nd}/L$}  &
          \multicolumn{1}{c}{$U_4$}  & 
          \multicolumn{1}{c}{$U_6$} \\
\hline
\hline
4.0&0.6843895(20)[15]&0.6843887(21)[14]&0.6843898(31)[20]&0.6843898(31)[26] \\
5.0&0.6875638(21)[16]&0.6875633(26)[15]&0.6875655(34)[20]&0.6875646(34)[26] \\
\end{tabular}
\end{table}

Then, we locate $\lambda^*$ by studying the scaling corrections 
to a quantity $\bar{R}$ defined in terms of two dimensionless ratios
$R_1$ and $R_2$. 
To define $\bar{R}$, we fix a number $R_{1,f}$ which 
should be a good approximation to $R_1^*$, see Ref. \cite{CHPRV-01}.
Then, for a given value of $\lambda$ and $L$, 
we determine $\beta_f(L,\lambda)$ from
\begin{equation}
\label{defbetaf}
R_1(L,\lambda,\beta_f) = R_{1,f} \; .
\end{equation}
In our analysis, 
$\beta_f$ is determined by taking either $(Z_a/Z_{p})_f=0.1944$ or 
$(\xi_{\rm 2nd}/L)_f=0.5644$. 
Note that $\beta_f$ approaches $\beta_c$ as 
\begin{equation}
\beta_f = \beta_c + C_f L^{-1/\nu} + ..., 
\end{equation}
where the prefactor $C_f$ depends on the choice of $R_{1,f}$.
In particular, if $R_{1,f}=R_{1}^*$, then $C_f=0$ 
and the leading corrections are proportional to $L^{-1/\nu-\omega}$.

Next, we define $\bar{R}$  by
\begin{equation}
\bar{R}(L,\lambda)\equiv R_2(L,\lambda,\beta_f) \;.
\end{equation}
Here, we take either
$U_4$ or $U_6$ as $R_2$. Below, we often refer to $\bar{R}$ as 
$R_2$ at $R_{1,f}$.
Up to subleading corrections, $\bar{R}$ behaves as
\begin{equation}
\label{corran}
\bar{R}(L,\lambda)\approx \bar{R}^* + \bar{c}(\lambda) L^{-\omega} \;\;.
\end{equation}
The optimal value $\lambda^*$ is obtained by solving $\bar{c}(\lambda) = 0$.
We obtain $\lambda^*=4.6(4)$, $4.7(8)$, $4.7(8)$ and $4.6(8)$  from 
$U_4$ at $(Z_a/Z_p)_f=0.1944$,
$U_4$ at $(\xi_{\rm 2nd}/L)_f=0.5644$,
$U_6$ at $(Z_a/Z_p)_f=0.1944$ and
$U_6$ at $(\xi_{\rm 2nd}/L)_f=0.5644$, respectively.
As our final result we quote 
\begin{equation} \lambda^*=4.6(4)
\end{equation}
  from $U_4$ at 
$(Z_a/Z_p)_f=0.1944$.

Finally, we  compute the critical exponents $\nu$ and $\eta$ using standard
FSS methods. 
Usually, the exponent $\nu$ is computed from
the slope of a dimensionless ratio $R$ at $\beta_c$.
Here, following Ref. \cite{BFMM-96}, we replace 
$\beta_c$ by $\beta_f$, which simplifies the error analysis,
and determine $\nu$ from the relation
\begin{equation}
\label{nuansatz}
 \left . \frac{\partial R}{\partial \beta} \right |_{\beta_f} = 
\bar{a} \; L^{1/\nu} .
\end{equation}
We study the derivative of all four quantities $U_4$, $U_6$,
$\xi_{\rm 2nd}/L$, and $Z_a/Z_p$, and fix $\beta_f$ by using either  
$(\xi_{\rm 2nd}/L)_f=0.5644$ or $(Z_a/Z_p)_f=0.1944$.
We arrive at the final estimate 
\begin{equation}
\nu=0.7113(11), 
\end{equation}
where 
the error includes both the statistical and the systematic
uncertainty.

The exponent $\eta$ is computed from the finite-size behavior of the 
magnetic susceptibility:
\begin{equation}
\label{chifit}
 \left . \chi \right |_{\beta_f} = c L^{2-\eta} \;.
\end{equation}
In addition,
we also use a fit ansatz that includes a constant background term:
\begin{equation}
\label{chiback}
 \left . \chi \right |_{\beta_f} = c L^{2-\eta} + b \;.
\end{equation}
As before, we  fix 
$\beta_f$ by setting either $(\xi_{\rm 2nd}/L)_f=0.5644$ or 
$(Z_a/Z_p)_f=0.1944$. Our final MC estimate is
\begin{equation}
\eta = 0.0378(6).
\end{equation}

\section{Critical exponents from the improved high-temperature expansion}
\label{HTanalysis}

As shown in the case of the Ising \cite{CPRV-99}
and $XY$ universality classes \cite{CPRV-00,CHPRV-01},
the analysis of HT expansions for improved Hamiltonians with
suppressed leading scaling corrections leads to considerably precise 
results even for moderately long series.
In the present paper, the analysis of  20th-order
HT expansions for the improved $\phi^4$ lattice Hamiltonian, i.e. for
$\lambda\approx \lambda^*=4.6(4)$, allows us to substantially improve the
accuracy of the estimates of the critical exponents. 
As we shall see, the results turn out to be more precise than those obtained 
in the preceding Section.  
They also significantly improve
those obtained from the analysis of longer series (21 orders)
for the standard Heisenberg model (which is recovered in the 
limit $\lambda\rightarrow\infty$)
on the cubic and bcc lattices \cite{BC-97}. 
In this Section we report the results of our analyses of the HT
series.  The details are reported in App.\ \ref{seriesanalysis}.

We determine $\gamma$ and $\nu$ from the analysis of the HT expansion
to $O(\beta^{20})$ of the magnetic susceptibility and of the
second-moment correlation length.  In App.\ \ref{crexpHT} we report
some details and  intermediate results so that the reader can judge
the quality of our results without the need of performing his own
analysis.  They should give an idea of the reliability of our
estimates and of the meaning of the errors we quote, which depend on
many somewhat arbitrary choices and are therefore partially
subjective.

We analyze the HT series by means of integral approximants (IA's) of
first, second, and third order.  The most precise results are obtained
biasing the value of $\beta_c$ with its MC estimate.  We consider
several sets of biased IA's and for each of them we obtain
estimates of the critical exponents.  These results are reported in
App.\ \ref{crexpHT}.  All sets of IA's give substantially consistent
results. Moreover, the results are also stable with respect to the
number of terms of the series, so that there is no need to perform
problematic extrapolations in the number of terms in order to obtain
the final estimates.  The error due to the uncertainty on $\lambda^*$
is estimated by considering the variation of the results
when changing the values of $\lambda$.

Using the results reported in App.\ \ref{crexpHT} 
for the analysis at $\lambda=4.0,4.5$, and $5.0$,  we obtain
\begin{eqnarray}
&&\gamma = 1.39582(10)[18] + 0.0015 (\lambda - 4.5),\label{gaphi4}\\
&&\nu    = 0.71111(5)[8]   + 0.0009 (\lambda - 4.5).\label{nuphi4}
\end{eqnarray}
The number between parentheses is basically the spread of
the approximants at $\lambda=4.5$ using the central value of
$\beta_c$, while the number between brackets gives the systematic
error due to the uncertainty on $\beta_c$. Eqs.\ (\ref{gaphi4}) and
(\ref{nuphi4}) show also the dependence of the results on the chosen
value of $\lambda$.  
The $\lambda$-dependence is estimated by using the results for 
$\lambda=4.0$ and $\lambda=5.0$.

Using the MC estimate $\lambda^*=4.6(4)$, we obtain
\begin{eqnarray}
&&\gamma = 1.39597(10)[18]\{60\},\\
&&\nu    = 0.71120(5)[8]\{36\},
\label{phi4exp}
\end{eqnarray}
where the error due to the uncertainty on $\lambda^*$ is reported
between braces. Thus, our final estimates are 
\begin{eqnarray}
&&\gamma=1.3960(9),\\
&&\nu=0.7112(5),
\end{eqnarray}
where the uncertainty is estimated by summing the
three errors reported above.

Using the above-reported results for $\gamma$ and $\nu$ and 
the scaling relation  $\gamma=(2-\eta)\nu$,
we obtain  $\eta=0.037(2)$,
where the error is estimated by considering the errors on $\gamma$ and
$\nu$ as independent, which is of course not true.  We can obtain an
estimate of $\eta$ with a smaller, yet reliable, error by applying the
so-called critical-point renormalization method  \cite{CPRM} 
to the series of $\chi$ and $\xi^2$.
This method provides an estimate for the combination $\eta\nu$.
Proceeding as before, we obtain
\begin{equation}
\eta\nu = 0.02665(18) + 0.00035 (\lambda-4.5). 
\end{equation}
Taking into account that $\lambda^*=4.6(4)$, we find
\begin{equation}
\eta\nu = 0.02669(18)[14],
\end{equation}
where the first error is related to the spread of the IA's and the
second one to the uncertainty on $\lambda^*$, evaluated as before.
Thus, 
\begin{equation}
\eta=0.0375(3)[2].
\end{equation}
Moreover, using the scaling relations, one obtains
\begin{eqnarray}
\alpha&=&2 - 3\nu=  -0.1336(15), \label{alphaex}\\ 
\delta &=& {5- \eta\over 1 +\eta}= 4.783(3), \label{deltaex}\\
\beta &=& {\nu\over 2} \left( 1 + \eta\right) = 0.3689(3), \label{betaex}
\end{eqnarray}
where the error of $\beta$ has been estimated by considering the errors of
$\nu$ and $\eta$ as independent.

\section{The critical equation of state}
\label{CES}

In this section we determine the critical equation of state 
characterizing the   Heisenberg universality class. 
The critical equation of state relates the 
thermodynamical quantities in the neighborhood of the critical
temperature, in both phases.  
It is usually written in the form 
\begin{eqnarray}
&&\vec{H} = (B_c)^{-\delta}\vec{M} M^{\delta-1} f(x), \\
&&x \equiv t (M/B)^{-1/\beta},
\label{eqstfx}
\end{eqnarray}
where $f(x)$ is a universal scaling function
normalized in such a way that $f(-1)=0$ and $f(0)=1$, 
and $B_c$ and $B$ are the amplitudes of the magnetization on the critical 
isotherm and on the coexistence curve,
\begin{eqnarray}
    M &=& B_c H^{1/\delta}\qquad\qquad t=0, \label{def-Bcconst} \\
    M &=& B (-t)^\beta\qquad\qquad  H=0,\,\, t<0. \label{def-Bconst}
\end{eqnarray}
Griffiths' analyticity implies that $f(x)$ is regular everywhere for $x>-1$. 
It has a regular expansion in powers of $x$,
\begin{equation}
f(x) = 1 + \sum_{n=1}^\infty f_n^0 x^n,
\label{expansionfx-xeq0}
\end{equation}
and a large-$x$ expansion of the form
\begin{equation}
f(x) = x^\gamma \sum_{n=0}^\infty f_n^\infty x^{-2n\beta}.
\label{largexfx}
\end{equation}
Moreover, at the coexistence curve, i.e. for $x\rightarrow -1$
\cite{BW-73,BZ-76,WZ-75,SH-78,Lawrie-81}
\begin{equation}
f(x) \approx  c_f \,(1+x)^2. 
\label{fxcc} 
\end{equation}
The nature of the corrections to the leading behavior at the
coexistence curve is less
clear, see, e.g., Refs.~\cite{WZ-75,SH-78,Lawrie-81,PV-99,PV-review}.
From the scaling function $f(x)$ one may derive 
many interesting  universal amplitude ratios involving 
zero-momentum quantities, such as 
specific heat, magnetic susceptibility, etc....
For example,
the universal ratio $U_0$ of the specific-heat amplitudes in the two phases
can be written as (see, e.g., Ref.~\cite{BHK-75})  
\begin{equation}
U_0\equiv {A^+\over A^-} = {\varphi(\infty)\over \varphi(-1)}
\end{equation}
where, in the Heisenberg case for which $-1<\alpha<0$,
\begin{equation}
\varphi(x) = 
{x |x|^{\alpha-2} f'(0) \over \alpha-1}  +
{|x|^{\alpha} f''(0)\over \alpha} - |x|^{\alpha-2} f(x) + 
\int_0^x d y\, |y|^{\alpha-2}
\left[ f'(y) - f'(0) - y f''(0) \right].
\label{U0fx}
\end{equation} 
We mention that the critical equation of state for the $N$-vector
model has been computed to $O(\epsilon^2)$ in the framework of the
$\epsilon$-expansion \cite{BWW-72}
and to $O(1/N)$ in the framework of the $1/N$ expansion \cite{BW-73}.

As our starting point for the determination 
of the critical equation of state, we determine
the first few nontrivial coefficients  
of its small-magnetization expansion, by analyzing 
the corresponding HT series for the improved $\phi^4$ Hamiltonian.
These results are then used  
to construct parametric representations of the
critical equation of state which are valid in the whole critical
region. Then, from our approximate representations of the
equation of state we derive estimates of several universal amplitude
ratios.  This method has been already applied to
the Ising universality class 
in three \cite{GZ-97,CPRV-99} and two dimensions \cite{CHPV-01}, 
and to the three-dimensional $XY$ universality class \cite{CPRV-00-2,CHPRV-01}.

\subsection{Small-magnetization expansion of the Helmholtz free energy}
\label{IHTrj}

We write the Helmholtz free energy as \cite{GZ-97,ZJbook}
\begin{equation}
\Delta {\cal F} = {\cal F}(M) - {\cal F}_{\rm reg}(M) = {m^3\over g_4}A(z),
\label{dAZ}
\end{equation}
where $m=1/\xi$, $\xi$ is the second-moment correlation length,
$g_4$ is the zero-momentum four-point coupling, 
and 
\begin{equation}
z\equiv k |M| t^{-\beta}
\end{equation}
where $k$ is an appropriate amplitude ratio.
The small-magnetization expansion of the free energy corresponds to
the small-$z$ expansion of $A(z)$,
\begin{equation}
A(z) = {1\over 2} z^2 + {1\over 4!} z^4
       + \sum_{j=3} {1\over (2j)!} r_{2j} z^{2j},
\label{AZ}
\end{equation}
which also fixes the normalization $k$ of $z$.
Correspondingly, we obtain for the equation of state
\begin{equation}
\vec{H} = {\partial {\cal F}(M)\over \partial \vec{M}} \propto 
{\vec{M}\over |M|} t^{\beta\delta} F(z),
\label{eqa}
\end{equation}
with $F(z) \equiv \partial A /\partial z$.
Because of Griffiths' analyticity, ${\cal F}(M)$  has a regular
expansion in powers of $t$ for $|M|$ fixed. Therefore,
$F(z)$ has the large-$z$ expansion
\begin{equation}
F(z) = z^\delta \sum_{k=0} F^{\infty}_k z^{-k/\beta}.
\label{asyFz}
\end{equation}
The function $F(z)$ is defined only for $t > 0$. For $t< 0$ the equation
of state is expressed in terms of a different function which is however
analytically related to $F(z)$ since the free energy and the equation of state
are analytic on the critical isotherm $t=0$ for $H\not=0$.
The two functions $f(x)$ and $F(z)$ are clearly related:
\begin{equation}
z^{-\delta} F(z) = F_0^\infty f(x), \qquad\qquad z = z_0 x^{-\beta},
\end{equation}
where $z_0 = k B$.

In order to estimate the universal quantities $g_4$ and $r_{2j}$ from
the corresponding improved HT expansions (see App.~\ref{HTexp}), 
we essentially used the analysis described in Ref.~\cite{CHPRV-01}.
Here, we report only the final estimates:
\begin{eqnarray}
&&g_4=19.13(8)[2], \\
&&r_6=1.86(3)[1],\\
&&r_8=0.60(15)[5],
\end{eqnarray}
where the error in parentheses is related to the spread of the 
approximants and the
second one in brackets to the uncertainty on $\lambda^*$, evaluated as before.
Moreover, we obtained the rough estimate $r_{10}= -15(10)$.
In Table \ref{summarygj} we compare  our results (denoted by IHT)
with the estimates obtained using other approaches, such as HT
expansions for the standard $O(3)$-vector 
model (HT), field-theoretical fixed-dimension
perturbative expansions ($d=3$ exp.), $\epsilon$ expansions
($\epsilon$-exp.), and approximate solutions of continuous 
renormalization-group equations (CRG).  
All estimates are in good agreement, Only the $\epsilon$-expansion estimate of 
$g_4$ is significantly higher than the IHT estimate
(as already noted in Ref.~\cite{PV-00}, the error may be
underestimated). 
The CRG estimates are much less precise than the results of other methods.

\begin{table}[tp]
\caption{\label{summarygj}
Estimates of $g_4$, $r_{6}$, and $r_8$, obtained using
various approaches.
}
\begin{tabular}{clllll}
\multicolumn{1}{c}{}&
\multicolumn{1}{c}{IHT}&
\multicolumn{1}{c}{HT}&
\multicolumn{1}{c}{$d=3$ exp.}&
\multicolumn{1}{c}{$\epsilon$-exp.}&
\multicolumn{1}{c}{CRG }\\
\tableline \hline
$g_4$   & 19.13(10)  & 19.31(14),$\;$ 19.27(11) \cite{BC-98}  & 19.06(5) \cite{GZ-98} &
          19.55(12) \cite{PV-00,PV-98-gr}  & 22.35 \cite{BTW-96,BTW-00}\\ 
        &    & 19.34(16) \cite{PV-98-gr}  &  19.06 \cite{MN-91} & & \\ \hline

$r_6$   & 1.86(4)  & 2.1(6) \cite{Reisz-95} & 1.880 \cite{SOUK-99} &
          1.867(9) \cite{PV-00,PV-98-ef} & 1.74 \cite{TW-94} \\
        & &  & 1.884(32) \cite{PV-00} & & \\ \hline

$r_8$   & 0.6(2)  & & 0.975 \cite{SOUK-99} &
          1.0(6) \cite{PV-00,PV-98-ef} & 0.84 \cite{TW-94}\\
\end{tabular}
\end{table}

\subsection{Parametric representations of the equation of state}
\label{preq}

In order to obtain approximations of the equation of state 
valid in the whole critical region, 
we use parametric representations that implement the expected
scaling and analytic properties. We write
\cite{Schofield-69,SLH-69,Josephson-69}
\begin{eqnarray}
M &=& m_0 R^\beta m(\theta) ,\nonumber \\
t &=& R(1-\theta^2), \nonumber \\
H &=& h_0 R^{\beta\delta}h(\theta), \label{parrep}
\end{eqnarray}
where $h_0$ and $m_0$ are normalization constants.  The variable $R$
is nonnegative and measures the distance from the critical point in
the $(t,H)$ plane, while the variable $\theta$ parametrizes the
displacement along the lines of constant $R$. The functions
$m(\theta)$ and $h(\theta)$ are odd and normalized so that 
$m(\theta)=\theta+O(\theta^3)$ and $h(\theta)=\theta+O(\theta^3)$.
The smallest positive zero of $h(\theta)$, which should satisfy
$\theta_0>1$, corresponds to the coexistence curve, i.e., to $T<T_c$
and $H\to 0$. 
The parametric representation satisfies the requirements of regularity
of the equation of state. Singularities can appear only at the
coexistence curve (due, for example, to the logarithms discussed in
Ref.\ \cite{PV-99}), i.e., for $\theta=\theta_0$.  
The mapping (\ref{parrep}) is not invertible when its Jacobian vanishes,
which occurs when
\begin{equation}
Y(\theta) \equiv (1-\theta^2)m'(\theta) + 2\beta\theta m(\theta)=0.
\label{Yfunc}
\end{equation}
Thus, parametric representations based on the mapping (\ref{parrep})
are acceptable only if $\theta_0<\theta_l$ where $\theta_l$ is the
smallest positive zero of the function $Y(\theta)$.  

The functions $m(\theta)$ and $h(\theta)$ are related to 
the scaling function $f(x)$ through
\begin{eqnarray}
&& x = {1 - \theta^2\over \theta_0^2 - 1} 
\left[ {m(\theta_0)\over m(\theta)}\right] ^{1/\beta}, \nonumber \\
&& f(x) = \left[ {m(\theta)\over m(1)}\right] ^{-\delta} {h(\theta)\over h(1)}.
\label{fxmt}
\end{eqnarray}
The asymptotic behavior (\ref{fxcc}) is reproduced simply by requiring that
\begin{equation}
h(\theta)\sim \left( \theta_0 - \theta\right)^2 
        \qquad\qquad{\rm for}\qquad \theta \rightarrow \theta_0.
\label{hcoex}
\end{equation}
The scaling function $F(z)$ is obtained by
\begin{eqnarray}
&&z = \rho \,m(\theta) \left( 1 - \theta^2\right)^{-\beta},
\nonumber \\
&&F(z(\theta)) = \rho \left( 1 - \theta^2 \right)^{-\beta\delta} h(\theta),
\label{Fzrel}
\end{eqnarray}
where $\rho$ may be taken as 
a free parameter \cite{GZ-97,CPRV-99,CPRV-00-2,PV-review}.  

\subsection{Approximate polynomial representations}
\label{appeq}

Following Ref.\ \cite{CPRV-00-2}, we construct approximate polynomial
parametric representations that have the expected singular behavior at
the coexistence curve (Goldstone singularity) and match the known 
terms of the small-$z$ expansion of $F(z)$, cf. Eqs. (\ref{AZ}) and (\ref{eqa}).
We consider two distinct approximation schemes.  In the first one,
which we denote by (A), $h(\theta)$ is a polynomial of fifth order
with a double zero at $\theta_0$, and $m(\theta)$ is a polynomial of
order $(1+2n)$:
\begin{eqnarray}
{\rm scheme}\quad({\rm A}):\qquad\qquad 
&&m(\theta) = \theta 
    \left(1 + \sum_{i=1}^n c_{i}\theta^{2i}\right), \nonumber \\
&&h(\theta) = \theta \left( 1 - \theta^2/\theta_0^2 \right)^2. 
\label{scheme1}
\end{eqnarray}
In the second scheme, denoted by (B), we set 
\begin{eqnarray}
{\rm scheme}\quad({\rm B}):\qquad\qquad 
&&m(\theta) = \theta, \nonumber \\
&&h(\theta) = \theta 
    \left(1 - \theta^2/\theta_0^2 \right)^2
    \left( 1 + \sum_{i=1}^n c_{i}\theta^{2i}\right).
\label{scheme2}
\end{eqnarray}
Here $h(\theta)$ is a polynomial of order $5+2n$ with a double zero at
$\theta_0$.  For $n=0$ the approximations (A) and (B) coincide.
Note that for scheme (B)
\begin{equation}
Y(\theta) = 1 - \theta^2 + 2\beta\theta^2,
\label{Ytheta_def}
\end{equation}
independently of $n$, so that $\theta_l = (1-2\beta)^{-1}$.
In both schemes, $\rho$,
$\theta_0$, and the $n$ coefficients $c_{i}$ are
determined by matching the small-$z$ expansion of $F(z)$. 
Thus, in order to fix the $n$ coefficients $c_i$ we
use $n+1$ values of $r_{2j}$, i.e., $r_6,...r_{6+2n}$.

\begin{figure}[tb]
\hspace{-1cm}
\vspace{0cm}
\centerline{\psfig{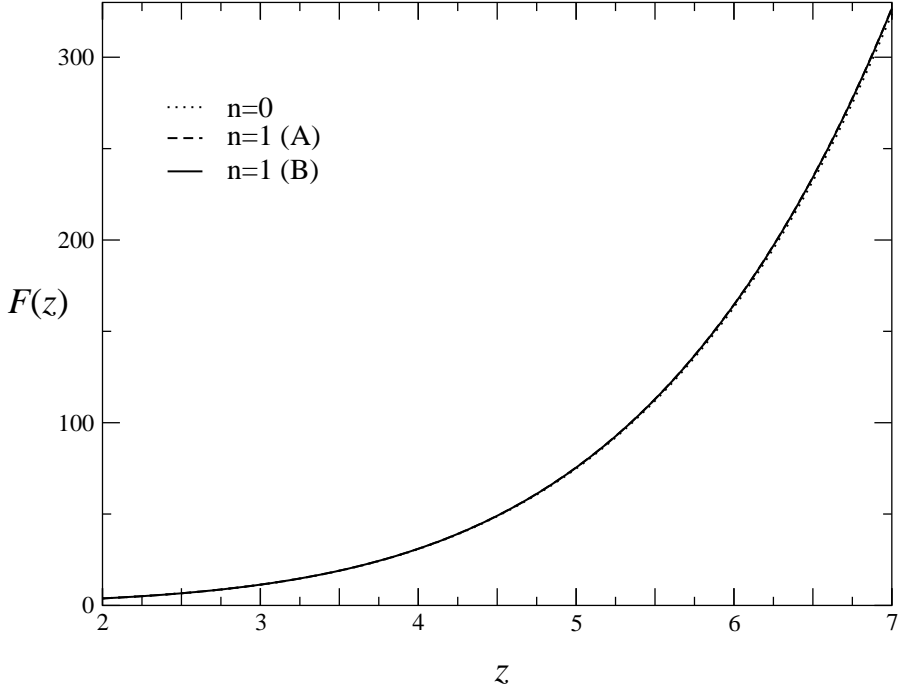}}
\vspace{0cm}
\caption{
The scaling function $F(z)$ versus $z$.}
\label{figFz}
\end{figure}

As input parameters for our analysis we consider the estimates 
$\alpha = -0.1336(15)$, $\eta=0.0375(5)$, $r_6=1.86(4)$, and
$r_8=0.6(2)$, which are the results of our HT
analysis. The available estimate of $r_{10}$ is too imprecise
for our purposes.
In Fig.\ \ref{figFz} we show the curves obtained in
schemes (A) and (B) with $n=0,1$ and for 
$\alpha = -0.1336$, $\eta=0.0375$, $r_6=1.86$, and $r_8=0.6$.
The differences among the three approximations should give
an indication of the uncertainty.
The three approximations of $F(z)$ are practically
indistinguishable, and differ at most by approximately 2\%
(the difference between the two $n=1$ curves is much smaller).
Thus, by using the first two coefficients $r_{2j}$, one
obtains reasonably precise approximations of the scaling function
$F(z)$ for all positive values of $z$, i.e., for the whole HT phase up
to $t=0$.  
This is also numerically confirmed
by the estimates of the universal constant $F_0^\infty$,
cf. Eq.~(\ref{asyFz}), which is related to the large-$z$ behavior of
$F(z)$. Indeed, we obtain $F_0^\infty=0.0262(4),\; 0.0266(5),\; 0.0266(5)$ 
respectively for $n=0$, $n=1$ (A), $n=1$ (B),
where the reported errors refer only to the uncertainty of the
input parameters.
This fact is not trivial, since the small-$z$
expansion has a finite convergence radius \cite{foot-YL}.
Therefore, the determination of
$F(z)$ on the whole positive real axis from its small-$z$ expansion
requires an analytic continuation, which turns out to be effectively
performed by the approximate parametric representations we have
considered.

\begin{figure}[tb]
\hspace{-1cm}
\vspace{0.2cm}
\centerline{\psfig{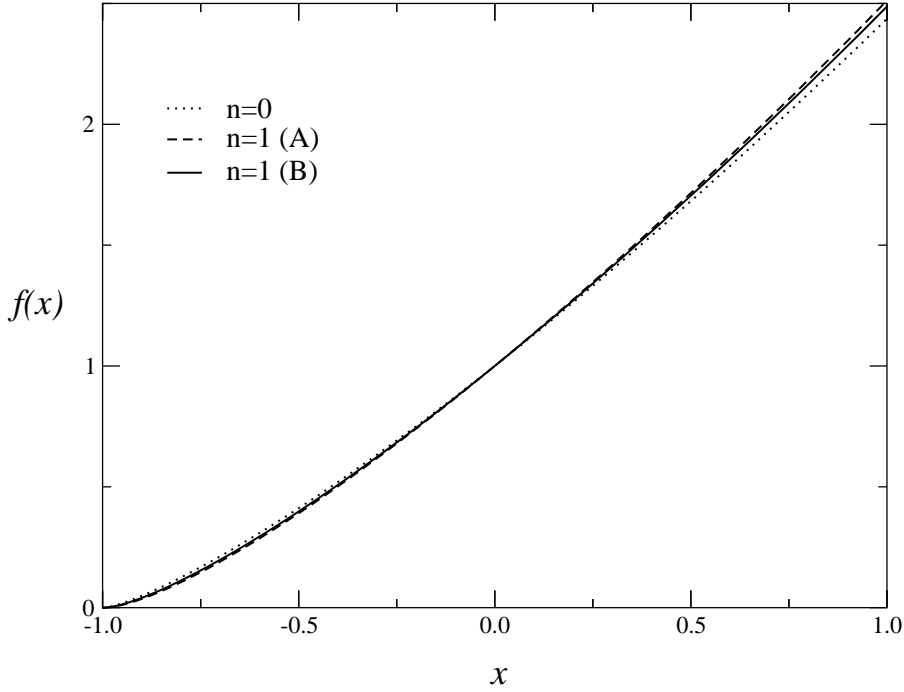}}
\vspace{-0.2cm}
\caption{
The scaling function $f(x)$ versus $x$.}
\label{figfx}
\end{figure}

In Fig.\ \ref{figfx} we plot the approximations of $f(x)$
corresponding to the schemes (A) and (B) for $n=0,1$, using the 
central values of the input parameters.
The three curves are in substantial agreement, especially 
those with $n=1$. Indeed, the difference between them 
is within the uncertainty due to the errors on the input parameters.
These approximate parametric representations are not 
precise at the coexistence curve; indeed, as we shall see, the
estimates of $c_f$, cf. Eq.~(\ref{fxcc}), 
are rather imprecise and very sensitive to the value of $r_8$,
which is not known with high precision. 
We mention that in Ref.~\cite{BTW-96} an  approximate expression for $f(x)$ was obtained
by approximately solving the continuous renormalization-group 
equations for the free energy
(average action). The results are quite imprecise, as we shall
show later by comparing the corresponding estimates for some universal
amplitude ratios.

\subsection{Universal amplitude ratios}
\label{unratio}

From the critical equation of state one may derive estimates
of several universal amplitude ratios.
They are expressed in terms of the amplitudes of the 
magnetization, cf. Eqs. (\ref{def-Bcconst}) and (\ref{def-Bconst}), 
of the singular part of the specific heat
\begin{equation}
C_{H,\rm sing} = A^{\pm} |t|^{-\alpha},
\end{equation}
of the magnetic susceptibility in the HT phase
\begin{equation}
\chi = N C^{+} t^{-\gamma},
\end{equation}
of the zero-momentum four-point connected correlation function in the
HT  phase
\begin{equation}
\chi_4 = {N(N+2)\over 3} C_4^+ t^{-\gamma-2\beta\delta}
\end{equation}
and of the second-moment correlation length in the HT phase
\begin{equation}
\xi = f^{+} t^{-\nu},
\end{equation}
where $N=3$.
We also consider the crossover (or pseudocritical) line $t_{\rm max}(H)$, 
that is defined as the reduced temperature for which
the longitudinal magnetic susceptibility $\chi_L(t,H)=\partial M / \partial H$ 
has a maximum at $H$ fixed. 
The renormalization group predicts 
\begin{eqnarray}
&&t_{\rm max}(H) = T_p H^{1/(\gamma+\beta)},\\
&&\chi_L(t_{\rm max},H)= C_p t_{\rm max}^{-\gamma}.
\end{eqnarray}
We consider several universal amplitude ratios:
\begin{eqnarray}
&& U_0 \equiv {A^+\over A^-}, \\
&& R_\chi \equiv {C^+ B^{\delta-1}\over (B_c)^\delta} ,\\
&& R_C \equiv {\alpha A^+ C^+\over B^2} ,\\
&& R_4 \equiv  - {C_4^+ B^2\over (C^+)^3}, \\
&& R_\xi^+ \equiv (\alpha A^+)^{1/3} f^+ = \left( {R_C R_4\over g_4}\right)^{1/3},\\
&& P_m \equiv { T_p^\beta B\over B_c},\\
&& P_c \equiv -{ T_p^{2\beta\delta} C^+\over C_4^+} = {P_m^{2\delta} \over R_\chi^2 R_4}, \\
&& R_p \equiv { C^+\over C_p}.  
\end{eqnarray}
Morever, we estimate
\begin{equation}
R_\alpha  \equiv {1 -  U_0\over \alpha},
\end{equation}
which, as suggested in Ref.~\cite{BHK-75}, should be less
sensitive to the value of $\alpha$ than $U_0$.
In App. \ref{univparrep} we give their expressions in terms of the 
functions $m(\theta)$ and $h(\theta)$.  

In Table \ref{eqstresAB} we report the universal amplitude ratios,
as derived by the approximate polynomial representations of the 
equation of state for $n=0,1$.
The reported errors are only due to the uncertainty of the
input parameters and do not include the systematic error of the
procedure, which may be determined by comparing the results of the
various approximations.  
In Table \ref{eqstresAB} 
we also show results for $z_{\rm max}$, $x_{\rm max}$
and $w_{\rm max}$
which are the values of the
scaling variable $z$, $x$ and $w$ ($w$ will be defined 
in Eq.~(\ref{Cw})) associated with the crossover line, 
$F_0^\infty$, cf. Eq.~(\ref{asyFz}), 
which is related to the large-$z$ behavior of $F(z)$,
$r_8$ and $r_{10}$, cf. Eqs.~(\ref{AZ}) and (\ref{eqa}),
which are  related to the small-$z$ expansion of $F(z)$,
$f_1^0$, $f_2^0$, and $f_3^0$, cf. Eq.~(\ref{expansionfx-xeq0}),
which are related to the expansion at $x=0$ of $f(x)$,
and $c_f$, cf. Eq.~(\ref{fxcc}),
which is related to the behavior at the coexistence curve.
Note that $f_0^\infty=R_\chi^{-1}$ where 
$f_0^\infty$ is related to the large-$x$ behavior of $f(x)$,
cf. Eq.~(\ref{largexfx}).
 
\begin{table}
\caption{
Results for the parameters and the
universal amplitude ratios  
using the scheme (A), cf. Eq.~(\protect\ref{scheme1}), and the scheme (B),
cf. Eq.~(\protect\ref{scheme2}).  
Note that the quantities reported in the first three lines
do not have a physical meaning, but are related to the
particular parametric representation employed.
Numbers marked with an asterisk are inputs, not predictions.  
}
\label{eqstresAB}
\begin{tabular}{lccc}
\multicolumn{1}{c}{}&
\multicolumn{1}{c}{$n=0$}&
\multicolumn{1}{c}{$n=1$$\;$ (A)}&
\multicolumn{1}{c}{$n=1$$\;$ (B)}\\
\tableline \hline
$\rho$       & 2.14(2)     & 2.20(4)     & 2.08(4)     \\
$\theta_0^2$ & 3.81(1)     & 3.3(1)      & 2.7(2)      \\
$c_1$        &  0          & $-$0.016(9) & 0.055(20) \\\hline
$U_0$    & 1.61(2)         & 1.56(3)     & 1.57(3) \\
$R_\alpha$    & 4.6(2)     & 4.2(3)      & 4.3(2) \\
$R_\chi$     & 1.41(2)     & 1.28(8)     & 1.33(4)  \\
$R_C$        & 0.173(3)    & 0.19(1)     & 0.184(6)  \\
$R_4$        & 8.2(2)      &  7.7(3)     & 7.9(2)    \\
$R_\xi^+$    & 0.421(1)    & 0.425(3)    & 0.423(2)  \\
$P_m$        & 1.201(5)    & 1.17(2)     & 1.18(1)  \\
$P_c$        & 0.354(4)    & 0.357(5)    & 0.357(5)  \\
$R_p$        & 2.026(6)    & 2.020(5)    & 2.021(7)  \\
$z_{\rm max}$ & 1.278(5)   & 1.275(5)    & 1.275(6)  \\
$x_{\rm max}$ & 8.9(1)     & 8.3(4)      & 8.5(2)  \\
$w_{\rm max}$ & 1.64(2)    & 1.53(6)     & 1.57(4)  \\
$F_0^{\infty}$& 0.0262(4)  & 0.0266(5)   & 0.0266(5) \\
$r_{8}$      & 0.23(5)     & $^*$0.6(2)  &$^*$0.6(2) \\
$r_{10}$     & $-$1.1(3)   & $-$6(2)     & $-$7(3) \\
$f_1^0$      & 1.28(1)     & 1.36(5)     &   1.33(3) \\
$f_2^0$      & 0.181(5)    & 0.21(2)     &    0.20(1)  \\
$f_3^0$      & $-$0.102(1) & $-$0.105(2) & $-$0.094(5)  \\
$c_f$        & 22(5)       & 5(3)        &  8(3)     \\
\end{tabular}
\end{table}

\begin{table}
\caption{
Estimates of universal amplitude ratios obtained using different approaches.
The numbers marked by an asterisk have been obtained by us using the
results reported in the corresponding references.
}
\label{univratios}
\begin{tabular}{cllllll}
\multicolumn{1}{c}{}&
\multicolumn{1}{c}{IHT--PR}&
\multicolumn{1}{c}{$d=3$ exp.}&
\multicolumn{1}{c}{$\epsilon$- exp.}&
\multicolumn{1}{c}{CRG}&
\multicolumn{1}{c}{HT}&
\multicolumn{1}{c}{experiments}\\
\tableline \hline
$U_0$ & 1.56(4) & 1.51(4) \cite{LMSD-98} & 1.521(22) \cite{Bervillier-86} & 
            $^*$1.823 \cite{BTW-96,BTW-00} & & 1.50(5) \cite{KR-94}\\
      &         & 1.544 \cite{KV-01} & & & &   1.27(9) \cite{MMFB-96} \\
      & & & & & & 1.4(4) \cite{Ramos-etal-01} \\ \hline
$R_\alpha$ & 4.3(3)  &  $^*$4.4(4) \cite{LMSD-98}  & 4.56(9)
           \cite{Bervillier-86} & $^*$3.41 \cite{BTW-96,BTW-00} &&  \\
           &         &  $^*$4.46 \cite{KV-01}  &   &&&  \\\hline

$R_\chi$ &  1.31(7) &  & 1.33 \cite{AM-78} & 1.11 \cite{BTW-96,BTW-00} & & \\\hline

$R_C$    & 0.185(10) & 0.189(9) \cite{SLD-99} & 0.17 \cite{AH-76} & & & \\
         & & 0.194 \cite{KV-01} & & & & \\\hline

$R_4$ & 7.8(3) & &&& & \\\hline

$R_\xi^+$ & 0.424(3) & 0.4347(20) \cite{BB-85} & 0.42 \cite{Bervillier-76} &&
0.431(5) \cite{BC-99} & \\
& & 0.4319(17) \cite{BG-80} & & & 0.433(5) \cite{BC-99}& \\ 
\end{tabular}
\end{table}

From the results of Table~\ref{eqstresAB} we arrive at the final
estimates denoted by IHT--PR in Table~\ref{univratios},
obtained by taking the weighted average
of the results for $n=1$. The error we quote 
is the sum of the uncertainty induced by the error on the input parameters
and of one half of the difference between the two approximations with $n=1$.
In most cases these estimates include the results of the  $n=0$
approximation. 
In Table~\ref{univratios} we compare our results 
with those obtained in  other approaches and in experiments 
\cite{foot-Rchi}. 
We mention that the field-theoretical estimates of $U_0$ 
have been obtained from the analysis
of the fixed-dimension expansion in the framework of the
minimal renormalization without $\epsilon$-expansion \cite{LMSD-98,KV-01},
and from the standard $\epsilon$-expansion to $O(\epsilon^2)$ \cite{Bervillier-86}.
The estimate of $U_0$ by CRG was  obtained using Eq.~(\ref{U0fx}) and the 
approximate expression for $f(x)$ reported in Refs.~\cite{BTW-96,BTW-00}.
See, e.g., Ref.~\cite{PHA-91} for a more complete review of 
theoretical and experimental estimates of universal amplitude ratios.

\begin{figure}[tb]
\hspace{-1cm}
\vspace{0cm}
\centerline{\psfig{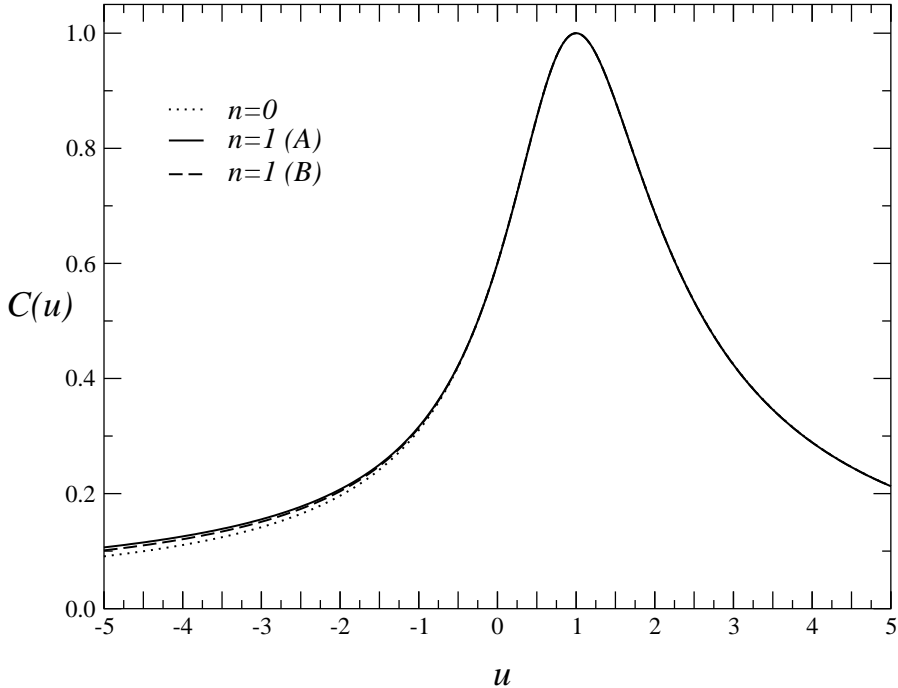}}
\vspace{0cm}
\caption{
The scaling function $C(u)$, cf. Eqs.~(\ref{Cwb}) and (\ref{Cw}).
}
\label{figchil}
\end{figure}

In addition, from the approximate parametric
representations of the equation of state, we obtain the estimates
\begin{eqnarray}
&&f_1^0=1.34(5), \nonumber \\
&&f_2^0=0.20(2), \nonumber \\
&&f_3^0=-0.10(1), \nonumber \\ 
&&F_0^\infty=0.0266(5), \nonumber \\
&&r_{10}=-6(3). \label{altre-stime-rapporti}
\end{eqnarray}
The estimate of $r_{10}$ 
should be compared with the much less precise HT result
$r_{10}=-15(10)$ obtained in Sec.~\ref{IHTrj}.
Concerning the quantities involving amplitudes at 
the crossover line, we report the estimates
\begin{eqnarray}
&&P_m = 1.18(2),\\
&&P_c = 0.357(5),\\
&& R_p = 2.020(6),\\
&&z_{\rm max} = 1.275(5).
\end{eqnarray}
In order to determine  the behavior of the longitudinal magnetic
susceptibility $\chi_L=\partial M/\partial H$ as a function of $t$ and
$H$, one may consider the scaling function 
\begin{eqnarray}
&&D(w) \equiv   B_c^{-1} H^{1-1/\delta} \chi_L =
 { f(x)^{1-1/\delta} \over \delta f(x) - \case{1}{\beta} x f'(x)},
\nonumber \\
&& w \equiv     (B/B_c)^{1/\beta} t H^{-1/(\beta\delta)} = 
x f(x)^{-1/(\beta\delta)}.
\label{Cw} 
\end{eqnarray}
The function 
$D(w)$ has a maximum for $w_{\rm max} = 1.55(6)$.
In order to simplify possible comparisons, it is convenient to
consider the rescaled function
\begin{eqnarray}
&&C(u) = {D(w)\over D(w_{\rm max}) }, \nonumber \\
&& u = {w\over w_{\rm max}}, \label{Cwb}
\end{eqnarray}
which is such that 
the maximum corresponds to $u=1$ and satisfies $C(1) =1$.
In Fig.~\ref{figchil}  we plot the scaling function $C(u)$ versus $u$, 
as obtained from the $n=0,1$ approximate parametric representations.

\subsection{Comparison with the experiments} \label{CES.E}

In spite of the large number of experiments, at present there is no 
accurate quantitative study of the equation of state in the critical 
regime. We shall discuss here three different representations that 
are widely used in the experimental work and we shall give explicit 
formulae for them. 

\begin{figure}[tb]
\hspace{-1cm}
\vspace{0cm}
\centerline{\psfig{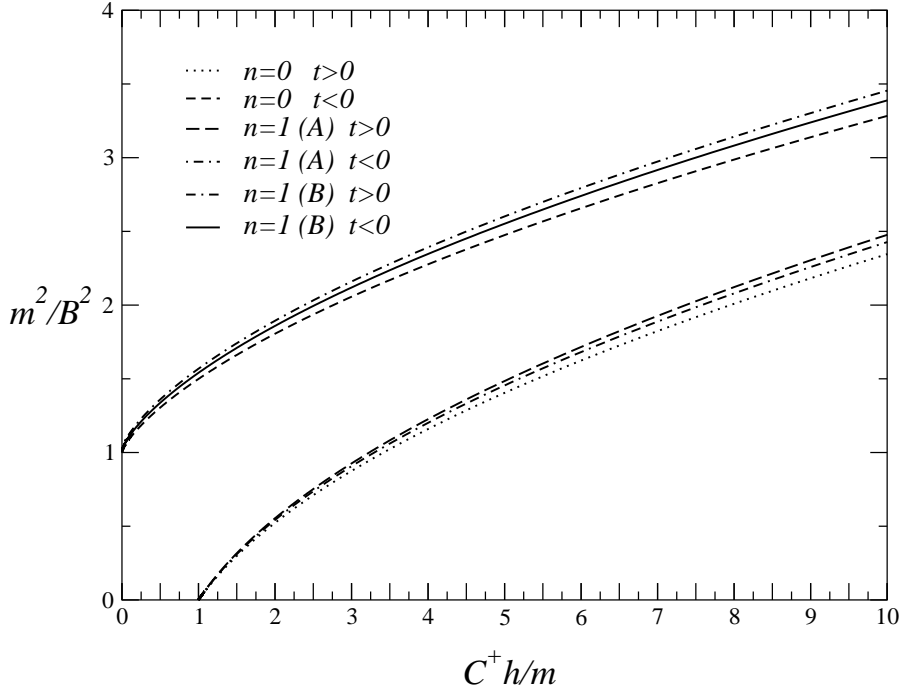}}
\vspace{0cm}
\caption{
Plot of $m^2/B^2$ versus $C^+ h/m$.
}
\label{m2vshsum}
\end{figure}

A first possibility \cite{KR-67} consists in studying the behavior of 
$h/m\equiv H |t|^{-\gamma}/M$ versus $m^2 = M^2 |t|^{-2\beta}$. Such a function 
can be easily obtained from our approximations for $f(x)$, since 
$m^2 = B^2 |x|^{-2\beta}$ and 
\begin{equation} 
   {h\over m} = k |x|^{-\gamma} f(x)
\end{equation}
where the constant $k$ can be written as 
\begin{equation}
   k = \left(B_c \right)^{-\delta} B^{\gamma/\beta} = {R_\chi\over C^+}.
\end{equation}
The universal ratio $R_\chi$ has been computed in the previous section, 
$R_\chi = 1.31(7)$, and $B$ and $B_c$ are nonuniversal amplitudes defined in 
Eqs. (\ref{def-Bconst}) and (\ref{def-Bcconst}).
A plot of $m^2/B^2$ versus $C^+ h/m$ is reported in Fig. \ref{m2vshsum}.
It agrees qualitatively with the analogous experimental ones 
reported, e.g., in Refs. \cite{KR-94,BK-97,FKK-01}.
Often, for small $h/m$ one approximates the equation of state by
writing 
\begin{equation}
   {h\over m} = a_{\pm} + b_{\pm} m^2,
\label{Kouvell-Rodbell}
\end{equation}
where $a_{\pm}$ and $b_\pm$ are numerical coefficients depending on the phase.
Such an approximation has a very limited range of validity. In the HT
phase, we obtain for $m^2 \to 0$
\begin{eqnarray}
{h\over m} &=& {1\over C^+} \left[ 1 + {R_4\over 6} {m^2\over B^2} + 
   \sum_{n=2}^\infty {R^n_4 \ r_{2n+2}\over (2n+1)!} 
   \left({m^2\over B^2}\right)^n \right]
\nonumber \\
&\approx& {1\over C^+} \left[ 1 + 1.30(5)\ {m^2\over B^2} + 
              0.94(8)\ \left({m^2\over B^2}\right)^2 + 
              0.06(2)\ \left({m^2\over B^2}\right)^3 + \cdots\right],
\label{hsum-expansion}
\end{eqnarray}
where we have used the estimate of $R_4$ reported in Table \ref{univratios}
and the estimates of $r_6$ and $r_8$ reported in Table \ref{summarygj}. 
From Eq. (\ref{hsum-expansion}) we see that the 
approximation (\ref{Kouvell-Rodbell}) is valid only for very small $m^2$, 
i.e. at the 1\% level only for $m^2 \lesssim 0.01 B^2$. 
The quadratic approximation---i.e. the approximation with an 
additional $(m^2)^2$ term---has a much wider range of validity 
because of the smallness of the coefficient of $m^6$.

In the low-temperature phase, Eq. (\ref{Kouvell-Rodbell}) is theoretically
incorrect, since for $m^2/B^2\to 1$ we have 
\begin{equation}
{h\over m} \approx {k c_f\over 4 \beta^2} \left(1 - {m^2\over B^2}\right)^2,
\label{hsum-coex}
\end{equation}
where $c_f$ is defined in Eq. (\ref{fxcc}) and can be estimated roughly from
the results reported in Table \ref{eqstresAB}.
Eq. (\ref{hsum-coex}) is inconsistent with the approximation
(\ref{Kouvell-Rodbell}) near the coexistence curve,
due to the presence of Goldstone modes. It would be correct only 
in Ising systems.

Finally, note that for $m^2$ large 
we have
\begin{equation}
{h\over m} \approx k \left({m\over B}\right)^{\delta-1}.
\end{equation}
A second form that is widely used to analyze the experimental data 
is the Arrott-Noakes \cite{AN-67} scaling equation
\begin{equation}
\left({H\over M}\right)^{1/\gamma} = a t + b M^{1/\beta},
\end{equation}
where $a$ and $b$ are numerical constants. This approximation 
is good in a neighborhood of the critical isotherm $t=0$. 
Since
\begin{equation}
\left({H\over M}\right)^{1/\gamma} k^{-1/\gamma} = 
\left({M\over B}\right)^{1/\beta} f(x)^{1/\gamma},
\end{equation}
using Eq. (\ref{expansionfx-xeq0}) and the numerical values reported in 
Eq. (\ref{altre-stime-rapporti}), we obtain
\begin{equation}
\left({H\over M}\right)^{1/\gamma} k^{-1/\gamma} = 
\left({M\over B}\right)^{1/\beta} + 
 0.96(4)\ t - 0.04(2)\, t^2 \left({M\over B}\right)^{-1/\beta}
            - 0.02(2)\, t^3 \left({M\over B}\right)^{-2/\beta}
        \cdots 
\end{equation}
Thus, at a 1\% level of precision the Arrott-Noakes formula is valid approximately 
for $t (M B^{-1})^{-1/\beta} \lesssim  25$ which is quite a large interval.
  
\begin{figure}[tb]
\hspace{-1cm}
\vspace{0cm}
\centerline{\psfig{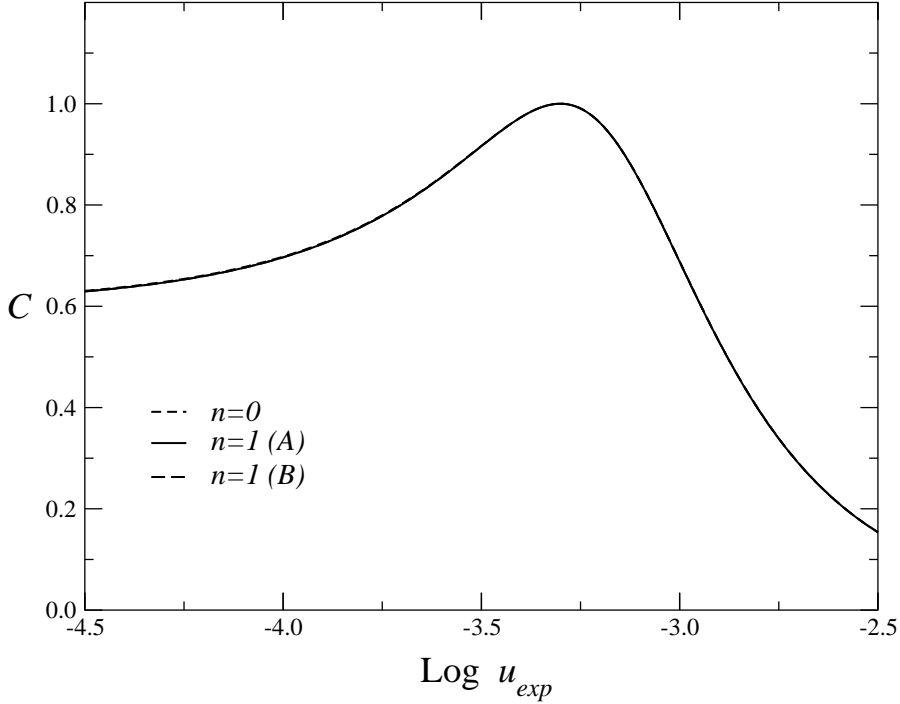}}
\vspace{0cm}
\caption{
The scaling function $C(u)$ versus the experimental scaling variable 
${u}_{\rm exp} \equiv t H^{-1/(\beta\delta)}$.
}
\label{figchilexp}
\end{figure}

Finally, Ref.~\cite{Zhao-etal-99} reports an experimental study of
the behavior of the critical system at the crossover line,
and shows a plot of the curve $C(u)$, cf. Eq.~(\ref{Cwb}), 
in terms of the unnormalized variable 
${u}_{\rm exp} \equiv t H^{-1/(\beta\delta)}$. 
We can attempt a quantitative comparison with
the results reported in their Fig.~4. 
For this purpose, in Fig.~\ref{figchilexp} we plot $C(u)$ 
in terms of their variable for the range of ${u}_{\rm exp}$ accessible to 
the experiment \cite{foot4}. A direct comparison of 
this figure with Fig.~4 of Ref.~\cite{Zhao-etal-99} shows a very nice 
quantitative agreement.

\appendix

\section{Monte Carlo Simulations}
\label{MonteCarlo-appendix}

In this appendix we present some details of the analyses of 
the MC data. Details on the simulation can be found in 
Ref.~\cite{Hasenbusch-00}. 

\subsection{Definitions}

In all our work, considerable importance is played by 
dimensionless ratios (or phenomenological couplings) $R$. 
In order to have better
control on systematic errors we have studied four dimensionless ratios.
We first consider the  Binder cumulant $U_4$ 
\cite{binder} and its generalization $U_6$ defined by
\begin{equation}
 U_{2j} \;=\; \frac{\langle(\vec{m}^2)^j\rangle}{\langle\vec{m}^2\rangle^j} 
  \;\;\;,
\end{equation}
where
\begin{equation}
 \vec{m} \; =\; \frac{1}{V} \; \sum_x \; \vec{\phi}_x
\end{equation}
is the magnetization of the system. 
The third quantity that we studied is
the second-moment correlation length divided by the linear extension
of the lattice $\xi_{\rm 2nd}/L$.
The second-moment correlation length is defined by
\begin{equation}
\xi_{\rm 2nd} \;=\; \sqrt{\frac{\chi/F-1}{4 \; \sin(\pi/L)^2}} \;\;\;,
\end{equation}
where
\begin{equation}
\chi \;=\; \frac{1}{V} \;
 \left\langle \left(\sum_x \; \vec{\phi}_x \right)^2 \right\rangle
\end{equation}
is the magnetic susceptibility and
\begin{equation}
F \;= \; \frac{1}{V} \;   \left \langle
\left|\sum_x \exp\left(i \frac{2 \pi x_1}{L} \right) \vec{\phi}_x \right|^2 \;\;
\right \rangle
\end{equation}
is the Fourier transform of the two-point correlation function at
the lowest nonvanishing momentum. In order to reduce the statistical
error, we averaged the results of all three directions of the lattice.

The fourth quantity is the ratio $Z_a/Z_p$, where 
$Z_a$ is the partition function with anti-periodic boundary
conditions in one of the three directions and $Z_p$ the corresponding one 
with periodic boundary conditions in all directions. 
Anti-periodic boundary conditions mean that the term
$\sum_{<xy>} \vec{\phi}_x \cdot \vec{\phi}_y$ in the Hamiltonian
is multiplied by $-1$
for $x=(L_1,x_2,x_3)$ and $y=(1,x_2,x_3)$.
This ratio can be measured with the
help of a variant of the cluster algorithm, the boundary-flip algorithm.
It was introduced in Ref. \cite{Ha-93}
for the Ising model and generalized 
to $O(N)$-invariant nonlinear $\sigma$ models in Ref. \cite{GH-94}. 
As in Refs.~\cite{Hasenbusch-99,CHPRV-01},
we use a version of the algorithm that only measures $Z_a/Z_p$ and does not 
perform the flip to anti-periodic boundary conditions.
For a recent discussion of the algorithm, see Ref. \cite{CHPRV-01}.

\subsection{Determination of $R^*$}
\label{betacdet}

First, we compute 
$\beta_c$ and the fixed-point value of 
the dimensionless ratios $R^*$ for $\lambda=4.5$---our best approximation 
to $\lambda^*$---using the standard 
cumulant crossing method of Binder.

For $\lambda = 4.5$ we solve Eq. (\ref{bindercross}), computing 
$R(L,\beta)$ by using its Taylor expansion up to the third order:
\begin{equation}
R(L,\beta) = R(L,\beta_s) \;+ d_1(L,\beta_s) \; (\beta-\beta_s)
                   \;+ \; \smfrac{1}{2} \; d_2(L,\beta_s) (\beta-\beta_s)^2
     \;+ \; \smfrac{1}{6} \; d_3(L,\beta_s) (\beta-\beta_s)^3.
                                     \end{equation}
Here $\beta_s$ is the value of $\beta$ at which the simulation was performed,
and $R$, $d_1$, $d_2$, and $d_3$ are determined in
the MC simulation.

As an example, results for $Z_a/Z_p$ are given in Table \ref{betac}.
In the fits, we include all data with $L_{\rm min}\le L \le L_{\rm max}$. 
For $L_{\rm max}=96$, $\chi^2/$d.o.f. (d.o.f. is the number of degrees of 
freedom of the fit) is smaller than 1 starting from 
$L_{\rm min}=16$. Moreover, the result for $\beta_c$ is stable when further 
data are discarded.
To be on the safe side, we take our final estimate 
from the fit with $L_{\rm min}=28$ and $L_{\rm max}=96$. 

The systematic error due
to corrections to scaling is estimated by comparing the results 
corresponding to $L_{\rm min}=28$ and $L_{\rm max}=96$ with those with
$L_{\rm min}=14$ and $L_{\rm max}=48$.  We suppose that  the 
error of $\beta_c$ is proportional to $L^{-1/\nu-\omega} \approx L^{-2.2}$,
where we assume pessimistically leading and not subleading corrections.
Hence, we estimate the error on our final result  as the 
difference between the $L_{\rm min}=28$, $L_{\rm max}=96$ result and the 
$L_{\rm min}=14$, $L_{\rm max}=48$ result multiplied by 
$2^{-2.2}/(1-2^{-2.2})$. 
The systematic error of $(Z_a/Z_p)^*$ is estimated in a similar fashion, 
assuming that the error is proportional to $L^{-\omega}$. 

In the same way we analyze our data for the remaining three 
dimensionless ratios. Our results are reported in Table \ref{betacR}.
Note that the four results for $\beta_c$ are consistent within error bars.
The statistical error of $\beta_c$ obtained from $Z_a/Z_p$ and 
$\xi_{\rm 2nd}/L$ is
considerably smaller than that from $U_4$ and $U_6$. 
As our final estimate we take $\beta_c=0.6862385(20)$, which is consistent with 
all four results.

\begin{table}
\caption{\label{betac} Fits with ansatz~(\ref{bindercross})
of our data for $Z_a/Z_p$ at $\lambda=4.5$. We included all data
with $L_{\rm min} \le L \le L_{\rm max}$ in the fit. 
}
\begin{tabular}{lllll}
\multicolumn{1}{c}{$L_{\rm min}$} &
\multicolumn{1}{c}{$L_{\rm max}$} &
\multicolumn{1}{c}{$\chi^2/$d.o.f.} &
\multicolumn{1}{c}{$\beta_c$} &
\multicolumn{1}{c}{$(Z_a/Z_p)^*$} \\
\hline
\hline
12 & 96 & 5.75 & 0.6862428(6)  & 0.19408(3)  \\
14 & 96 & 1.90 & 0.6862413(7)  & 0.19419(3)  \\
16 & 96 & 1.29 & 0.6862406(7)  & 0.19424(4)  \\
18 & 96 & 0.62 & 0.6862400(7)  & 0.19430(4)  \\
20 & 96 & 0.68 & 0.6862400(8)  & 0.19430(5)  \\
22 & 96 & 0.75 & 0.6862400(8)  & 0.19430(6)  \\
24 & 96 & 0.81 & 0.6862397(10) & 0.19434(8)  \\
28 & 96 & 0.60 & 0.6862390(10) & 0.19443(10) \\
12 & 48 & 5.96 & 0.6862459(10) & 0.19399(4)  \\
14 & 48 & 1.92 & 0.6862432(10) & 0.19413(4)  \\
\end{tabular}
\end{table}

\subsection{Determination of $\lambda^*$}

In order to compute $\lambda^*$ we solve the equation $\bar{c}(\lambda) = 0$,
where $\bar{c}(\lambda)$ is defined in Eq. (\ref{corran}).
In practice, we replace $\bar{c}(\lambda)$ with its first-order 
Taylor expansion around $\lambda = 4.5$ and thus 
evaluate $\lambda^*$ from 
\begin{equation}
\label{convert}
 \lambda^* \approx 4.5 - {\bar{c}(4.5)}
\left({\left. \frac{\mbox{d}\bar{c}}{\mbox{d} \lambda} \right|_{\lambda=4.5}}
  \right)^{-1} \;\;\;.
\end{equation}                                                                  
In order to compute $\bar{c}(4.5)$ we fit
our data for $\bar{R}$ with the ansatz (\ref{corran}),
where we fix $\omega=0.8$.  We have checked that the 
final result for $\lambda^*$ has a very weak dependence on the value of 
$\omega$ used in the analysis. 
If we choose $\omega = 0.75$, the results vary much less than
the quoted error bar, indicating that the error 
on $\omega$ can be neglected. 

As an example, the results for $U_4$ at $(Z_a/Z_p)_f=0.1944$ are given 
in Table \ref{corrzazpu}. 
We see that there is a slight drift towards larger values of $\bar{c}(4.5)$
as $L_{\rm min}$ is increased. 
The final result corresponds to $L_{\rm min}=16$ and $L_{\rm max}=96$.
Systematic errors due to subleading corrections to scaling 
are estimated by comparing with the results obtained for 
$L_{\rm min}=8$ and $L_{\rm max}=48$.
Assuming the errors on $\bar{c}(4.5)$ to decrease as 
$L^{-\omega_2+\omega} \approx L^{-0.8}$, 
we arrive at $\bar{c}(4.5)=0.0010(6)[19]$, where the systematic 
error is quoted in brackets.

\begin{table}
\caption{\label{corrzazpu}  
Estimates of $\bar{R}^*$ and $\bar{c}(4.5)$ from the fit (\ref{corran}) of 
$U_4$ at $(Z_a/Z_p)_f=0.1944$. Here $\lambda = 4.5$. 
}
\begin{tabular}{rllll}
\multicolumn{1}{c}{$L_{\rm min}$} & 
\multicolumn{1}{c}{$L_{\rm max}$} & 
\multicolumn{1}{c}{$\chi^2/$d.o.f.}  &
\multicolumn{1}{c}{$\bar{R}^*$}    &
\multicolumn{1}{c}{$\bar{c}(4.5)$} \\
\hline
\hline
  6  &  96   &   7.87  &   1.13931(2)    &\phantom{+}0.00073(12) \\
  8  &  96   &   1.71  &   1.13944(2)    &$-$0.00036(17) \\
 10  &  96   &   1.82  &   1.13942(3)    &$-$0.00022(25) \\
 12  &  96   &   1.65  &   1.13937(4)    &\phantom{+}0.00039(36) \\
 14  &  96   &   1.25  &   1.13930(5)    &\phantom{+}0.00118(47) \\
 16  &  96   &   1.33  &   1.13932(5)    &\phantom{+}0.00097(60) \\
 18  &  96   &   1.33  &   1.13928(6)    &\phantom{+}0.00145(73) \\
 20  &  96   &   1.27  &   1.13923(7)    &\phantom{+}0.00217(89) \\
 24  &  96   &   1.54  &   1.13920(10)   &\phantom{+}0.00273(155) \\
  8  &  32   &   1.00  &   1.13947(3)    &$-$0.00060(19) \\
  8  &  48   &   1.26  &   1.13945(2)    &$-$0.00045(18) \\
 12  &  48   &   1.20  &   1.13939(4)    &\phantom{+}0.00018(38) \\
 16  &  48   &   0.75  &   1.13936(6)    &\phantom{+}0.00056(67) \\
\end{tabular}
\end{table}

In Table \ref{corrxi2u} we give the analogous results for $U_4$ at 
$(\xi_{\rm 2nd}/L)_f=0.5644$.
\begin{table}
\caption{\label{corrxi2u}  
Estimates of $\bar{R}^*$ and $\bar{c}(4.5)$ from the fit (\ref{corran}) of 
$U_4$ at $(\xi_{\rm 2nd}/L)_f=0.5644$. Here $\lambda = 4.5$. 
}
\begin{tabular}{rlrll}
\multicolumn{1}{c}{$L_{\rm min}$} & 
\multicolumn{1}{c}{$L_{\rm max}$} & 
\multicolumn{1}{c}{$\chi^2/$d.o.f.}  &
\multicolumn{1}{c}{$\bar{R}^*$}    &
\multicolumn{1}{c}{$\bar{c}(4.5)$} \\
\hline
\hline
  8  &  96  &   9.98   &  1.13984(2) & $-$0.00493(19) \\
 10  &  96  &   5.05   &  1.13966(3) & $-$0.00313(31) \\
 12  &  96  &   2.32   &  1.13948(4) & $-$0.00120(39) \\
 14  &  96  &   1.30   &  1.13937(5) & \phantom{+}0.00010(51) \\
 16  &  96  &   1.40   &  1.13936(6) & \phantom{+}0.00022(65) \\
 18  &  96  &   1.21   &  1.13929(7) & \phantom{+}0.00112(81) \\
 20  &  96  &   1.24   &  1.13925(8) & \phantom{+}0.00167(99) \\
 24  &  96  &   1.50   &  1.13923(11)& \phantom{+}0.00209(173)\\
  8  &  48  &  10.41   &  1.13987(3) & $-$0.00514(19) \\
 12  &  48  &   1.99   &  1.13952(4) & $-$0.00153(42) \\
\end{tabular}
\end{table}
Here, we see a larger change of $\bar{c}(4.5)$ 
when $L_{\rm min}$ is varied. Also,  $\chi^2/$d.o.f. is larger 
for $L_{\rm min} < 14$. 
Since corrections are larger than above, 
we take the final estimate from the fit with 
$L_{\rm min}=24$ and $L_{\rm max}=96$.
We arrive at the estimate $\bar{c}(4.5)=0.002(2)[5]$. 
In a similar way we arrive at 
     $\bar{c}(4.5)=0.007(5)[15]$ for $U_6$ at $(Z_a/Z_p)_f=0.1944$ 
and  $\bar{c}(4.5)=0.003(4)[18]$ for $U_6$ at $(\xi_{\rm 2nd}/L)_f=0.5644$.

Next, we compute $\mbox{d}\bar{c}/\mbox{d} \lambda$ at 
$\lambda=4.5$.
To estimate the derivative of $\bar{c}(\lambda)$,
we consider the finite differences
\begin{equation}
\left . \frac{\mbox{d}\bar{c}}{\mbox{d} \lambda} \right|_{\lambda=4.5} 
\approx \left[\bar{R}(L,5.0)-\bar{R}(L,4.0)\right]
\; L^{\omega} \;\;.
\end{equation}
The results for our 
four choices of $\bar{R}$ are given in Table \ref{derivative}.
We see that the results, as functions of $L$, are constant within error bars.
This nicely confirms  the exponent $\omega=0.8$. 

\begin{table}
\caption{\label{derivative} 
Estimates of $\left[\bar{R}(L,5.0)-\bar{R}(L,4.0)\right]\, L^{\omega}$
with  $\omega=0.8$.
}
\begin{tabular}{rllll}
\multicolumn{1}{c}{$L$} & 
\multicolumn{1}{c}{$U_4$ at $(Z_a/Z_p)_f$} &
\multicolumn{1}{c}{$U_4$ at $(\xi_{\rm 2nd}/L)_f$} & 
\multicolumn{1}{c}{$U_6$ at $(Z_a/Z_p)_f$} &
\multicolumn{1}{c}{$U_6$ at $(\xi_{\rm 2nd}/L)_f$} \\
\hline
\hline
  6  &$-$0.00957(17)   &  $-$0.01128(18) & $-$0.03141(53)  &$-$0.03674(54) \\
  8  &$-$0.00951(21)   &  $-$0.01136(22) & $-$0.03111(67)  &$-$0.03697(70) \\
 10  &$-$0.00973(25)   &  $-$0.01174(27) & $-$0.03164(80)  &$-$0.03799(85) \\
 12  &$-$0.00972(28)   &  $-$0.01160(30) & $-$0.03150(88)  &$-$0.03747(93) \\
 14  &$-$0.00983(33)   &  $-$0.01188(36) & $-$0.03201(105) &$-$0.03855(113)\\
 16  &$-$0.00972(37)   &  $-$0.01181(40) & $-$0.03183(115) &$-$0.03848(123)\\
 18  &$-$0.00972(47)   &  $-$0.01195(53) & $-$0.03177(148) &$-$0.03884(162)\\
 20  &$-$0.01022(57)   &  $-$0.01174(63) & $-$0.03392(179) &$-$0.03880(194)\\
 22  &$-$0.00947(68)   &  $-$0.01176(74) & $-$0.03074(213) &$-$0.03795(231)\\
 24  &$-$0.00962(79)   &  $-$0.01209(86) & $-$0.03128(249) &$-$0.03915(272)\\
 28  &$-$0.00962(124)  &  $-$0.01159(135)& $-$0.03208(388) &$-$0.03832(423) \\
\end{tabular}
\end{table}

The final result for the derivative is obtained by averaging the 
results for $L\ge12$, see Table \ref{avderiv}.  In order to 
estimate  the discretization error, we additionally compute  
the derivative, using $\bar{R}(L,\lambda)$
for the pair $\lambda=5.0$ and $\lambda=4.5$.
The difference with the above-reported result is small and in practice 
negligible, 
approximately  $12 \%$ for $U_4$ at $(Z_a/Z_p)_f=0.1944$ and 
$U_4$ at $(\xi_{\rm 2nd}/L)_f=0.5644$,
and approximately $14 \%$ for $U_6$ at $(Z_a/Z_p)_f=0.1944$
and $U_6$ at $(\xi_{\rm 2nd}/L)_f=0.5644$.

\begin{table}
\caption{\label{avderiv} Results for 
$\left[\bar{R}(L,5.0)-\bar{R}(L,\lambda)\right]/(5.0 - \lambda) \, L^{\omega}$ 
with
$\omega=0.8$ for $L\ge 12$. In the first row $\lambda = 4.0$, in the second one
$\lambda = 4.5$.
}
\begin{tabular}{llll}
\multicolumn{1}{c}{$U_4$ at $(Z_a/Z_p)_f$} &
\multicolumn{1}{c}{$U_4$ at $(\xi_{\rm 2nd}/L)_f$} & 
\multicolumn{1}{c}{$U_6$ at $(Z_a/Z_p)_f$} &
\multicolumn{1}{c}{$U_6$ at $(\xi_{\rm 2nd}/L)_f$} \\
\hline
\hline
$-$0.00976(16)  & $-$0.01177(17) & $-$0.03184(49) & $-$0.03823(53)  \\
$-$0.00872(27)  & $-$0.01035(29) & $-$0.02838(86) &  $-$0.03356(92)  \\
\end{tabular}
\end{table}

Inserting our numerical results for $\bar{c}(4.5)$ and 
$\mbox{d}\bar{c}/\mbox{d} \lambda$ into Eq.~(\ref{convert}) we get
$\lambda^*=4.6(4)$, $4.7(8)$, $4.7(8)$, and $4.6(8)$  from 
$U_4$ at $(Z_a/Z_p)_f=0.1944$,
$U_4$ at $(\xi_{\rm 2nd}/L)_f=0.5644$,
$U_6$ at $(Z_a/Z_p)_f=0.1944$ and
$U_6$ at $(\xi_{\rm 2nd}/L)_f=0.5644$, respectively.
The errors take into account the uncertainty of
$\bar{c}(4.5)$, 
$\left . \frac{\mbox{d}\displaystyle{\bar{c}}}{
               \mbox{d}\displaystyle{\lambda}} \right|_{\lambda=4.5}$, and 
$\omega$.

As our final result we quote $\lambda^*=4.6(4)$ from $U_4$ at 
$(Z_a/Z_p)_f=0.1944$.

%

\subsection{Critical exponents}
\label{exponentsMC}

We compute the critical exponents $\nu$ and $\eta$ using standard
FSS methods. 

\subsubsection{The exponent $\nu$}

The exponent $\nu$ is determined by fitting the data with 
Eq. (\ref{nuansatz}).
We study the derivative of all four quantities $U_4$, $U_6$,
$\xi_{\rm 2nd}/L$, and $Z_a/Z_p$, and fix $\beta_f$ by using either  
$(\xi_{\rm 2nd}/L)_f=0.5644$ or $(Z_a/Z_p)_f=0.1944$.

As typical examples, we give fit results for
$\left .\frac{\partial U_4}{\partial \beta} \right|_{\beta_f}$,
$\left .\frac{\partial (Z_a/Z_p)}{\partial \beta} \right|_{\beta_f}$
and
$\left .\frac{\partial (\xi_{\rm 2nd}/L)}{\partial \beta} \right|_{\beta_f}$
in Tables \ref{dUatZaZp}, \ref{dZaZpatZaZp}, and \ref{dxi2atZaZp}, 
respectively. In all these three cases, we have fixed 
$\beta_f$ by $(Z_a/Z_p)_f$. Fixing  $\beta_f$ by 
$(\xi_{\rm 2nd}/L)_f$ leads to similar results.

In the case of $\left .\frac{\partial U_4}{\partial \beta} \right|_{\beta_f}$
we see that  $\chi^2/$d.o.f. is close to one,
even if all lattice sizes $L \ge 6$ are included in the fit. 
Also, the result for $\nu$ stays rather stable when 
$L_{\rm min}$ is increased.

On the other hand, for 
$\left .\frac{\partial (Z_a/Z_p)}{\partial \beta} \right|_{\beta_f}$
and
$\left .\frac{\partial (\xi_{\rm 2nd}/L)}{\partial \beta} \right|_{\beta_f}$,
$\chi^2/$d.o.f.
comes close to one only starting from $L_{\rm min}  \ge 16$. Also, the results
for $\nu$ strongly change, when $L_{\rm min}$ is increased. It is interesting 
to notice that the estimate of $\nu$ is decreasing for 
$\left .\frac{\partial (\xi_{\rm 2nd}/L)}{\partial \beta} \right|_{\beta_f}$
while it is increasing for
$\left .\frac{\partial (Z_a/Z_p)}{\partial \beta} \right|_{\beta_f}$
when $L_{\rm min}$ is increased. Assuming that this is already the asymptotic
behavior, we can find lower and upper bounds for $\nu$.

\begin{table}
\caption{\label{dUatZaZp}
Estimates of $\nu$ from the fit of 
$\left .\frac{\partial U_4}{\partial \beta} \right|_{\beta_f}$  with 
ansatz~(\ref{nuansatz}). $\beta_f$ is fixed by $(Z_a/Z_p)_f=0.1944$.
}
\begin{tabular}{rlll}
\multicolumn{1}{c}{$L_{\rm min}$} &
\multicolumn{1}{c}{$L_{\rm max}$} & 
\multicolumn{1}{c}{$\chi^2/$d.o.f.} & 
\multicolumn{1}{c}{$\nu$} \\
\hline
\hline
  6 & 96 &  1.36 &  0.71215(16) \\
  8 & 96 &  0.72 &  0.71174(20) \\
 10 & 96 &  0.70 &  0.71179(26) \\
 12 & 96 &  0.71 &  0.71160(34) \\
 16 & 96 &  0.79 &  0.71137(49) \\
 20 & 96 &  0.86 &  0.71167(66) \\
\end{tabular}
\end{table}

\begin{table}
\caption{\label{dZaZpatZaZp}
Estimates of $\nu$ from the fit of 
$\left .\frac{\partial (Z_a/Z_p)}{\partial \beta} \right|_{\beta_f}$  with 
the ansatz~(\ref{nuansatz}). $\beta_f$ is fixed by $(Z_a/Z_p)_f=0.1944$.
}
\begin{tabular}{rlrl}
\multicolumn{1}{c}{$L_{\rm min}$} &
\multicolumn{1}{c}{$L_{\rm max}$} & 
\multicolumn{1}{c}{$\chi^2/$d.o.f.} & 
\multicolumn{1}{c}{$\nu$} \\
\hline
\hline
  6 & 96 & 69.37 &  0.70660(6) \\
  8 & 96 & 11.95 &  0.70837(8) \\
 10 & 96 &  3.28 &  0.70918(11)\\
 12 & 96 &  1.91 &  0.70969(15)\\
 16 & 96 &  1.40 &  0.71009(22)\\
 20 & 96 &  1.21 &  0.71054(30)\\
 24 & 96 &  1.46 &  0.71044(46)\\
 28 & 96 &  1.59 &  0.71071(58)\\
\end{tabular}
\end{table}

\begin{table}
\caption{\label{dxi2atZaZp}
Estimates of $\nu$ from the fit of 
$\left .\frac{\partial (\xi_{\rm 2nd}/L)}{\partial \beta} \right|_{\beta_f}$ 
with ansatz~(\ref{nuansatz}). $\beta_f$ is fixed by $(Z_a/Z_p)_f=0.1944$.
}
\begin{tabular}{rrrl}
\multicolumn{1}{c}{$L_{\rm min}$} &
\multicolumn{1}{c}{$L_{\rm max}$} & 
\multicolumn{1}{c}{$\chi^2/$d.o.f.} & 
\multicolumn{1}{c}{$\nu$} \\
\hline
\hline
 6 & 96 &68.10 & 0.71826(8)  \\
 8 & 96 &18.76 & 0.71613(11) \\
10 & 96 & 7.72 & 0.71489(14) \\
12 & 96 & 4.35 & 0.71394(18) \\
16 & 96 & 1.32 & 0.71261(27) \\
20 & 96 & 1.43 & 0.71246(37) \\
24 & 96 & 1.27 & 0.71169(56) \\
28 & 96 & 1.39 & 0.71196(70) \\
\end{tabular}
\end{table}

Taking into account the fit results for $L_{\rm min} \ge 22$ 
we arrive at the final estimate 
$\nu=0.7113(10)$.
Here, the error bar includes both the statistical and the systematic
error.


Finally, we try to determine the effect of leading corrections to scaling on 
our estimate of $\nu$.
For this purpose we fit our data up to $L_{\rm max}=28$
at $\lambda=4.0$, $\lambda=4.5$, and $\lambda=5.0$ 
with the ansatz~(\ref{nuansatz}).  In Table \ref{checknub} we give our results 
for the derivative of the Binder cumulant with respect to $\beta$ at 
$(Z_a/Z_p)_f=0.1944$. In particular, for small $L_{\rm min}$, we see a clear 
dependence of the result for $\nu$ on $\lambda$.
For instance, for 
$L_{\rm min}=8$ the difference between the result for $\lambda=4.0$ and 
$\lambda=5.0$ is $0.00176(46)$.

\begin{table}
\caption{\label{checknub}
Estimates of  $\nu$ computed from the derivative of the Binder cumulant
at $\beta_f$, where $\beta_f$ is fixed by $(Z_a/Z_p)_f=0.1944$. 
}
\begin{tabular}{rlrl}
\multicolumn{1}{c}{$L_{\rm min}$} &
\multicolumn{1}{c}{$L_{\rm max}$} &
\multicolumn{1}{c}{$\chi^2/$d.o.f.} & 
\multicolumn{1}{c}{$\nu$} \\
\hline
\hline
\multicolumn{4}{c}{$\lambda=4.0$}  \\
\hline
\hline
 6 & 28 &  0.91 &  0.71301(24) \\
 8 & 28 &  0.75 &  0.71269(33) \\
10 & 28 &  0.83 &  0.71296(44) \\
12 & 28 &  1.07 &  0.71275(64) \\
\hline
\hline
\multicolumn{4}{c}{$\lambda=4.5$}\\
\hline
\hline
 6 & 28 &  1.64 &  0.71226(17) \\
 8 & 28 &  0.74 &  0.71181(23) \\
10 & 28 &  0.67 &  0.71193(32) \\
12 & 28 &  0.74 &  0.71170(46) \\
\hline
\hline
\multicolumn{4}{c}{$\lambda=5.0$}\\
\hline
\hline
 6 & 28 & 1.27  & 0.71153(23) \\
 8 & 28 & 0.76  & 0.71093(32) \\
10 & 28 & 0.44  & 0.71034(43) \\
12 & 28 & 0.51  & 0.71014(60) \\
\end{tabular}
\end{table}

In Table \ref{checknuz} we give the corresponding analysis for 
$Z_a/Z_p$ at $(Z_a/Z_p)_f=0.1944$. In this case we see a much smaller 
dependence of the results for $\nu$ on $\lambda$. For $L_{\rm min}=8$
the difference between the results for $\lambda=4.0$ and $\lambda=5.0$ is 
$-0.00065(18)$. The behavior in the case of $\xi_{\rm 2nd}/L$ (which is not 
shown here) is much the same: the dependence of the fit result for $\nu$ 
on $\lambda$ is much smaller than for the Binder cumulant.

Taking into account the range the lattice sizes that are used to obtain 
the final result for $\nu$, we arrive at a possible
uncertainty of $0.0001$ for $\nu$ from the derivative of $Z_a/Z_p$ and of 
$\xi_{\rm 2nd}/L$  due to the uncertainty in $\lambda^*$. The systematic
error is clearly dominated by subleading corrections.
Our final MC estimate of $\nu$ is
\begin{equation}
\nu=0.7113(11). 
\end{equation}

\begin{table}
\caption{\label{checknuz}
Estimates of  $\nu$ computed from the derivative of $Z_a/Z_p$
at $\beta_f$, where $\beta_f$ is fixed by $(Z_a/Z_p)_f=0.1944$. 
}
\begin{tabular}{rrrl}
\multicolumn{1}{c}{$L_{\rm min}$} &
\multicolumn{1}{c}{$L_{\rm max}$} &
\multicolumn{1}{c}{$\chi^2/$d.o.f.} & 
\multicolumn{1}{c}{$\nu$} \\
\hline
\hline
\multicolumn{4}{c}{$\lambda=4.0$}  \\
\hline
\hline
 8 & 28 &  8.97 &  0.70766(13) \\
10 & 28 &  1.63 &  0.70871(18) \\
12 & 28 &  1.58 &  0.70902(27) \\
\hline
\hline
\multicolumn{4}{c}{$\lambda=4.5$}\\
\hline
\hline
  8 & 28 & 14.38 &  0.70804(9)  \\
 10 & 28 &  2.84 &  0.70890(13) \\
 12 & 28 &  1.57 &  0.70941(19) \\
\hline
\hline
\multicolumn{4}{c}{$\lambda=5.0$}\\
\hline
\hline
  8 & 28 &  8.52  &  0.70831(13) \\
 10 & 28 &  2.89  &  0.70916(18) \\
 12 & 28 &  2.40  &  0.70965(25) \\
\end{tabular}
\end{table}

\subsubsection{The exponent $\eta$}

We determine the exponent $\eta$ by using Eq. (\ref{chifit})
and also a fit ansatz that includes a constant background term, 
Eq. (\ref{chiback}).
We  fix $\beta_f$ by setting either $(\xi_{\rm 2nd}/L)_f=0.5644$ or 
$(Z_a/Z_p)_f=0.1944$.

Fits for $\lambda=4.5$ 
with the ansatz~(\ref{chifit}) are given in Table \ref{tablechiatz} 
($\beta_f$ fixed by $(Z_a/Z_p)_f=0.1944$) and 
\ref{tablechiatxi2} ($\beta_f$ fixed by $(\xi_{\rm 2nd}/L)_f=0.5644$). 
In both cases,  $\chi^2/$d.o.f. becomes close to one 
starting from $L_{\rm min}=24$. Moreover, in both cases the fit results for 
$\eta$ are strongly increasing as $L_{\rm min}$ is increased. 
For $L_{\rm min}=32$
we have a consistent result of $\eta=0.0374(2)$.

\begin{table}
\caption{\label{tablechiatz}
Estimates of $\eta$ from 
fits of the magnetic susceptibility at $\lambda=4.5$ 
with Eq.~(\ref{chifit}). $\beta_f$ is fixed by 
$(Z_a/Z_p)_f=0.1944$.
}
\begin{tabular}{llrl}
\multicolumn{1}{c}{$L_{\rm min}$} &
\multicolumn{1}{c}{$L_{\rm max}$} &
\multicolumn{1}{c}{$\chi^2/$d.o.f.} & 
\multicolumn{1}{c}{$\eta$} \\
\hline
\hline
12 & 96 & 32.55 &  0.03557(5) \\
16 & 96 & 6.36  &  0.03641(7) \\
20 & 96 & 1.73  &  0.03682(9) \\
24 & 96 & 1.00  &  0.03710(13) \\
28 & 96 & 0.81  &  0.03725(16) \\
32 & 96 & 0.81  &  0.03740(24) \\
\end{tabular}
\end{table}

\begin{table}
\caption{\label{tablechiatxi2}
Estimates of $\eta$ from 
fits of the magnetic susceptibility at $\lambda=4.5$ 
with Eq.~(\ref{chifit}).
$\beta_f$ is fixed by $(\xi_{\rm 2nd}/L)_f=0.5644$.
}
\begin{tabular}{llrl}
\multicolumn{1}{c}{$L_{\rm min}$} &
\multicolumn{1}{c}{$L_{\rm max}$} &
\multicolumn{1}{c}{$\chi^2/$d.o.f.} & 
\multicolumn{1}{c}{$\eta$} \\
\hline
\hline
12 & 96 & 33.81 & 0.03592(4) \\
16 & 96 &  8.24 & 0.03656(5) \\
20 & 96 &  2.35 & 0.03695(7) \\
24 & 96 &  0.77 & 0.03725(10)\\
28 & 96 &  0.59 & 0.03735(12)\\
32 & 96 &  0.65 & 0.03742(18)\\
\end{tabular}
\end{table}

Next, we checked the dependence of the result for $\eta$ on $\lambda$. 
In Table \ref{tablechiatz2} we give results for $\lambda=4.0$, $4.5$, and 
$5.0$ for $L_{\rm min}=16$ and $L_{\rm max}=28$.  We see a rather strong 
dependence 
on $\lambda$. The difference between the results for $\lambda=4.0$ and 
$5.0$ is $0.00080(25)$.
Taking into account the range of lattice sizes used to obtain
our final estimate, 
we arrive at an error of $0.0002$ on $\eta$ 
due to the error on $\lambda^*$.

\begin{table}
\caption{\label{tablechiatz2}
Estimates of $\eta$ from 
fits of the magnetic susceptibility at $\lambda=4.0,4.5,5.0$ 
with Eq.~(\ref{chifit}).
$\beta_f$ is fixed by $(Z_a/Z_p)_f=0.1944$.
}
\begin{tabular}{lllll}
\multicolumn{1}{c}{$\lambda$} &
\multicolumn{1}{c}{$L_{\rm min}$} &
\multicolumn{1}{c}{$L_{\rm max}$} &
\multicolumn{1}{c}{$\chi^2/$d.o.f.} & 
\multicolumn{1}{c}{$\eta$} \\
\hline
\hline
4.0 & 16 & 28 & 4.75  &  0.03610(18) \\
4.5 & 16 & 28 & 2.96  &  0.03562(13) \\
5.0 & 16 & 28 & 8.53  &  0.03530(18) \\
\end{tabular}
\end{table}

Finally, we performed fits with the ansatz~(\ref{chiback}). The results
are summarized in Tables \ref{tablechiatzback} and \ref{tablechiatxi2back}.
We observe that a $\chi^2/$d.o.f. close to one is already reached for 
$L_{\rm min}=10$. Moreover, 
the result for $\eta$ changes little with increasing $L_{\rm min}$. 
For $L_{\rm min}=16$ the results obtained by fixing $\beta_f$ by $Z_a/Z_p$
and $\xi_{\rm 2nd}/L$ agree.  Therefore, we give as our final result 
$\eta=0.0378(6)$. The error bar is such that it includes the result
of the fits with ansatz (\ref{chifit}). 

\begin{table}
\caption{\label{tablechiatzback}
Estimates of $\eta$ from 
fits of the magnetic susceptibility at $\lambda=4.5$ 
with Eq.~(\ref{chiback}).
$\beta_f$ is fixed by $(Z_a/Z_p)_f=0.1944$.
}
\begin{tabular}{rllll}
\multicolumn{1}{c}{$L_{\rm min}$} &
\multicolumn{1}{c}{$L_{\rm max}$} &
\multicolumn{1}{c}{$\chi^2/$d.o.f.} & 
\multicolumn{1}{c}{$\eta$} &
\multicolumn{1}{c}{$b$}\\
\hline
\hline
 8 & 96 & 2.18  & 0.03832(8)  &$-$0.657(9) \\
10 & 96 & 1.22  & 0.03811(10) &$-$0.617(17) \\
12 & 96 & 0.72  & 0.03790(12) &$-$0.555(26) \\
16 & 96 & 0.66  & 0.03782(17) &$-$0.528(54) \\
\end{tabular}
\end{table}

\begin{table}
\caption{\label{tablechiatxi2back}
Estimates of $\eta$ from 
fits of the magnetic susceptibility at $\lambda=4.5$ 
with Eq.~(\ref{chiback}).
$\beta_f$ is fixed by $(\xi_{\rm 2nd}/L)_f=0.5644$.
}
\begin{tabular}{rllll}
\multicolumn{1}{c}{$L_{\rm min}$} &
\multicolumn{1}{c}{$L_{\rm max}$} &
\multicolumn{1}{c}{$\chi^2/$d.o.f.} & 
\multicolumn{1}{c}{$\eta$} &
\multicolumn{1}{c}{$b$}\\
\hline
\hline
 8 & 96 & 0.80  & 0.03756(6) & $-$0.379(6) \\
10 & 96 & 0.68  & 0.03765(7) & $-$0.396(11) \\
12 & 96 & 0.64  & 0.03772(9) & $-$0.416(20) \\
16 & 96 & 0.53  & 0.03780(13)& $-$0.454(39) \\
\end{tabular}
\end{table}

\section{Analysis of the high-temperature expansions}
\label{seriesanalysis}

In this appendix we report a discussion of our HT analyses.
It should allow the reader to understand how we
determined our estimates and the reliability of the errors we report,
which are to some extent subjective. More details on the methods 
we use are reported in Ref.~\cite{CHPRV-01}.

\subsection{Definitions and HT series}
\label{HTexp}

We computed the HT expansion of several quantities for 
the $\phi^4$ lattice Hamiltonian 
(\ref{phi4Hamiltonian}) for generic values of $\lambda$ by using the
linked-cluster expansion technique. 
A general introduction to this technique can be
found in Refs.\ \cite{Wortis-74,LW-88,Campostrini-01}. 
We calculated the 20th-order HT expansion of  
the magnetic susceptibility
and  of the second moment of the two-point function,
\begin{equation}
\chi = \sum_x \langle \phi_{\alpha}(0) \phi_{\alpha}(x) \rangle,
\qquad \qquad m_2 = \sum_x x^2 \langle \phi_{\alpha}(0) \phi_{\alpha}(x) \rangle,
\label{chi}
\end{equation}
and therefore, the second-moment correlation length $\xi^2=m_2/(6\chi)$.
Moreover, we computed the HT expansion of the zero-momentum
connected $2j$-point Green's functions $\chi_{2j}$
\begin{equation}
\chi_{2j} = \sum_{x_2,...,x_{2j}}
    \langle \phi_{\alpha_1}(0) \phi_{\alpha_1}(x_2) ...
        \phi_{\alpha_j}(x_{2j-1}) \phi_{\alpha_j}(x_{2j})\rangle_c
\end{equation}
($\chi = \chi_2$). More precisely, we computed $\chi_4$ to 18th order,
$\chi_6$, $\chi_8$, and $\chi_{10}$ to 15th order.  
In Table \ref{HTexpansions} we report 
the series for the $\phi^4$ Hamiltonian with
$\lambda=4.5$. We chose this value because it is very close to the 
best estimate
of $\lambda^*$, and because for this value of $\lambda$ we have
a precise MC estimate of $\beta_c$,
$\beta_c=0.6862385(20)$.

\begin{table}[tbp]
\caption{Coefficients of the HT expansion of $m_2$, $\chi$,
$\chi_4$, $\chi_6$,, $\chi_8$, and $\chi_{10}$,
for the $\phi^4$ Hamiltonian with $\lambda = 4.5$.
}
\label{HTexpansions}
\begin{tabular}{cccc}
$i$& \multicolumn{1}{c}{$m_2$} &
     \multicolumn{1}{c}{$\chi_2$} &
     \multicolumn{1}{c}{$\chi_4$} \\
\hline   
0 &                          0  &  0.95784805390722532625540  &  
     $-$0.10220686631889066464185 \\ 
1 &  0.61164859624923922306340  &  1.83494578874771766919200  &  
     $-$0.78318918399604286311017 \\ 
2 &  2.34346567036967995931352  &  3.02570526555699314952643  &  
     $-$3.40993303251195614433724 \\ 
3 &  6.12631563804734885224064  &  4.91084272357011167693574  &  
     $-$11.4547261149387426111654 \\ 
4 &  13.7411678013148223984708  &  7.72413685583625911405950  &  
     $-$32.8622857609807070588938 \\ 
5 &  28.1806907736271607364126  &  12.0696651111374510782698  &  
     $-$84.8703770522664536068131 \\ 
6 &  54.6110112231156495979858  &  18.5898743639602203433897  &  
     $-$203.048391944861326183480 \\ 
7 &  101.601106321810555490226  &  28.5129469684875374524852  &  
     $-$458.487070991901596469076 \\ 
8 &  183.444896900967255568497  &  43.3910414400431056256283  &  
     $-$989.113871576954399902821 \\ 
9 &  323.515763329983708761073  &  65.8457097219033943862880  &  
     $-$2056.80132192600410402482 \\ 
10 & 560.008592676223985571196  &  99.4318415601856883686915  &  
     $-$4148.88336851312159356144 \\ 
11 & 954.596109677157386424652  &  149.842255361185424824490  &  
     $-$8158.09343927729005801740 \\ 
12 & 1606.62254117132464855356  &  225.053071205898843291857  &  
     $-$15696.1032586702224741180 \\ 
13 & 2674.82852285795544112124  &  337.491827618915172447234  &  
     $-$29637.6862646277928041257 \\ 
14 & 4412.16935517377929559254  &  504.872666999718714410906  &  
     $-$55053.5845813513443015904 \\ 
15 & 7219.36840082433737483629  &  754.353044416651842398902  &  
     $-$100803.117025457078176935 \\ 
16 & 11729.5598928466629681760  &  1125.02916338959766441444  &  
     $-$182227.804691700019121686 \\ 
17 & 18938.5632727811981324466  &  1676.21770934179961650320  &  
     $-$325689.103869085626575337 \\ 
18 & 30408.8505644977121915813  &  2493.83291987123696064787  &  
     $-$576156.574987690512078391 \\ 
19 & 48583.1353446096892196268  &  3707.29637073719901187394  &  
      \\ 
20 & 77271.9486733817666941548  &  5504.79157669035824056791  &  
      \\ 
\hline\hline
$i$& \multicolumn{1}{c}{$\chi_6$} &
     \multicolumn{1}{c}{$\chi_8$} &
     \multicolumn{1}{c}{$\chi_{10}$} \\
\hline     
0 &  0.168561977829196181908019  &  
     $-$0.61605090918722894152663  &  3.90323863425838698945752 \\ 
1 &  2.564255760481036314377282  &  
     $-$15.2300029871348773005564  &  141.589060417753517915167 \\ 
2 &  19.75833949883362663601629  &  
     $-$178.372137615950383513749  &  2318.78554766886373709723 \\ 
3 &  107.6153442827461300177955  &  
     $-$1404.03404775262637194221  &  24643.8854034121990204490 \\ 
4 &  470.3177686849044939915090  &  
     $-$8533.18422765821528578124  &  196815.472836825966956507 \\ 
5 &  1762.676494500452240334969  &  
     $-$43117.7286797414375624292  &  1278183.93414715336465143 \\ 
6 &  5892.468315770018895075220  &  
     $-$189462.996356270457135473  &  7085909.43897471702725255 \\ 
7 &  18026.10606034742588181350  &  
     $-$745826.772293316301527782  &  34639336.4985891651938076 \\ 
8 &  51364.40898933529517266700  &  
     $-$2685696.43890603392620130  &  152792638.179196536855095 \\ 
9 &  138079.9439575146969655288  &  
     $-$8982884.52144712077544905  &  618543549.844101982857471 \\ 
10 & 353553.4243627006337358070  &  
     $-$28231531.3374765513159336  &  2328085189.97290506149579 \\ 
11 & 868613.6815956677724696610  &  
     $-$84122578.3791382746491922  &  8229819042.53291083435060 \\ 
12 & 2059460.213048375196521201  &  
     $-$239356596.354404154637789  &  27546203927.2418206714112 \\ 
13 & 4734189.661963741454010320  &  
     $-$654084831.403386579980857  &  87876514492.4881461446469 \\ 
14 & 10591072.53372872753915110  &  
     $-$1724768015.94141791734955  &  268647512699.595417711632 \\ 
15 & 23130642.47447337362164738  &  
     $-$4405965912.50799258526090  &  790611696518.867100914318 \\ 
\end{tabular}
\end{table} 

The HT series of the zero-momentum four-point coupling $g_4$ and of the
coefficients $r_{2j}$ that parametrize the small-magnetization
expansion of the equation of state can be
computed using their definitions in terms of $\chi_{2j}$ and $\xi^2$,
i.e.,
\begin{equation}
g_4 = - {3N\over N+2} {\chi_4\over \chi^2 \xi^3},
\label{grdef}
\end{equation}
and
\begin{eqnarray}
r_6 =&& 10 - {5(N+2)\over 3(N+4)}{\chi_6\chi_2\over \chi_4^2}, 
    \label{r2jgreen}\\
r_8 =&& 280 - {280 (N+2)\over 3(N+4)}{\chi_6\chi_2\over \chi_4^2} 
+{35(N+2)^2\over 9(N+4)(N+6)}{\chi_8\chi_2^2\over \chi_4^3}, \nonumber\\
r_{10} =&& 
15400  
-{7700  (N + 2)\over (N + 4)} {\chi_6 \chi_2\over \chi_4^2}        
+{ 350  (N + 2)^2\over(N + 4)^2} {\chi_6^2 \chi_2^2\over \chi_4^4} 
\nonumber \\ 
 &&+\;{1400 (N + 2)^2\over 3(N + 4)(N + 6)} {\chi_8\chi_2^2\over\chi_4^3}
    - {35 (N + 2)^3\over 3(N + 4) (N + 6) (N + 8)} 
                   {\chi_{10} \chi_2^3\over \chi_4^4}.
\nonumber
\end{eqnarray}
The formulae relevant for the Heisenberg universality class are obtained
setting $N=3$.

\subsection{Critical exponents}
\label{crexpHT}

In order to estimate $\gamma$ and $\nu$, we analyzed 
the 20th-order HT expansion of the magnetic susceptibility and the
19th-order HT expansion of $\xi^2/\beta$.
We analyzed the HT series by means of integral approximants
\cite{Guttrev} (IA's) of first, second, and third order 
(IA1's, IA2's and IA3's respectively).
Since the most precise results
are obtained by using the MC estimates of $\beta_c$ to bias the
approximants, we shall only report the results of the biased analyses.
We used the values of $\beta_c$ obtained in Sec.~\ref{betacdet}, i.e.
\begin{eqnarray}
&&\beta_c(\lambda=4.0)=0.6843895(35), \\
&&\beta_c(\lambda=4.5)=0.6862385(20),\\
&&\beta_c(\lambda=5.0)=0.6875638(37).
\end{eqnarray}
We considered several sets of biased IA's, and for each of them we obtained
estimates of the critical exponents.  In the analysis
we followed closely Ref.~\cite{CHPRV-01}. Thus, in the following
we shall heavily refer to it for notations and a more detailed
description of the analyses.

Given an $n$th-order series $f(\beta)= \sum_{i=0}^n c_i \beta^i$, its
$k$th-order integral approximant $[m_k/m_{k-1}/.../m_0/l]$ IA$k$ is a
solution of the inhomogeneous $k$th-order linear differential equation
\begin{equation}
P_k(\beta) f^{(k)}(\beta) + P_{k-1}(\beta) f^{(k-1)}(\beta) + ... 
+ P_1(\beta)f^{(1)}(\beta)+ P_0(\beta)f(\beta)+R(\beta)= 0,
\label{IAkdef}
\end{equation}
where the functions $P_i(\beta)$ and $R(\beta)$ are polynomials of
order $m_i$ and $l$ respectively, which are determined by the known
$n$th-order small-$\beta$ expansion of $f(\beta)$.
We considered two types of biased IA$k$'s:

(i) The first type of biased IA$k$'s, which will be denoted by
bIA$k$'s, is obtained by setting
\begin{equation}
P_k(\beta) = \left( 1 - \beta/\beta_c \right) p_k(\beta),
\end{equation}
where $p_k(\beta)$ is a polynomial of order $m_k-1$. 

(ii) Since on bipartite lattices $\beta=-\beta_c$ is also a singular
point associated to the antiferromagnetic critical behavior
\cite{Fisher-62}, we consider IA$k$'s with
\begin{equation}
P_k(\beta) = \left( 1 - \beta^2/\beta_c^2 \right) p_k(\beta),
\end{equation}
where $p_k(\beta)$ is a polynomial of order $m_k-2$.  We shall denote
them by b$_{\pm}$IA$k$'s.

In our analyses we considered diagonal or quasi-diagonal approximants,
since they are expected to give the most accurate results.  
For each set of IA$k$'s we calculated the average of the values
corresponding to all nondefective IA$k$'s. Approximants
are considered defective when they have singularities close to the
real $\beta$ axis near the critical point.  
We also discarded
some nondefective IA's---we call them outliers---whose results are far
from the average of the other approximants. 
All details can be found in the App.~B of Ref.~\cite{CHPRV-01}.

\begin{table}[tp]
\caption{\label{gamma}
Results for $\gamma$ obtained from the analysis of the 20th-order HT
series of $\chi$. 
}
\begin{tabular}{clcr@{}l}
\multicolumn{1}{c}{$\lambda$}&
\multicolumn{1}{c}{approximants}&
\multicolumn{1}{c}{$r_a$}&
\multicolumn{2}{c}{$\gamma$}\\
\tableline \hline
4.0
& bIA1$_{}$ & $ (35-3)/48$        & 1&.39508(6)[32]  \\

& bIA2$_{}$ & $ (77-7)/115$      & 1&.39503(16)[33] \\
\hline

4.5
& bIA1$_{}$ &     $ (36-3)/48$        & 1&.39585(4)[18]  \\

& b$_\pm$IA1$_{}$ & $ (21-1)/48$        & 1&.39583(4)[18]  \\

& bIA2$_{}$     & $ (93-11)/115$       & 1&.39580(10)[18] \\

& b$_\pm$IA2$_{}$     & $ (84-7)/100$       & 1&.39579(18)[18] \\

& bIA3$_{}$     & $ (56-6)/61$       & 1&.39582(5)[19] \\
\hline

5.0
& bIA1$_{}$ & $ (34-3)/48$        & 1&.39652(6)[32]  \\

& bIA2$_{}$ & $ (107-13)/115$     & 1&.39648(7)[34] \\
\end{tabular}
\end{table}

\begin{table}[tp]
\caption{\label{nu}
Results for $\nu$ obtained from the analysis of the 19th-order 
series of $\xi^2/\beta$. 
}
\begin{tabular}{clcr@{}l}
\multicolumn{1}{c}{$\lambda$}&
\multicolumn{1}{c}{approximants}&
\multicolumn{1}{c}{$r_a$}&
\multicolumn{2}{c}{$\nu$}\\
\tableline \hline
4.0
  & bIA1$_{}$ & $(37-6)/37$ & 0&.71061(1)[14] \\
  & bIA2$_{}$ & $(65-3)/70$ & 0&.71055(17)[14]   \\
\hline

4.5
  & bIA1$_{}$ & $(37-4)/37$ & 0&.71110(3)[8] \\

  & b$_\pm$IA1$_{}$ & $(31-3)/36$ & 0&.71111(2)[8] \\

  & bIA2$_{}$ & $(67-2)/70$ & 0&.71108(6)[7]   \\

  & b$_\pm$IA2$_{}$ & $(54-3)/55$ & 0&.71114(3)[8]   \\

  & bIA3$_{}$ & $(26-2)/34$ & 0&.71110(10)[10]   \\
\hline

5.0
  & bIA1$_{}$ & $(36-4)/37$ & 0&.71151(5)[15] \\
  & bIA2$_{}$ & $(67-5)/70$ & 0&.71154(6)[14]   \\
\end{tabular}
\end{table}

In Tables ~\ref{gamma} and \ref{nu} we report the results
for $\gamma$ and $\nu$ respectively, 
obtained by analyzing the series for $\lambda=4.0,4.5$, and $5.0$.
There, we also quote the ``approximant
ratio'' $r_a\equiv(g-f)/t$, where $t$ is the total number of
approximants in the given set, $g$ is the number of nondefective
approximants, and $f$ is the number of outliers which are discarded
using an algorithm described in App. B of Ref. \cite{CHPRV-01}; 
$g-f$ is the number of ``good''
approximants used in the analysis. Notice that $g \gg f$, and $g-f$ is
never too small.  For each analysis, beside the corresponding
estimate, we report two numbers.  The number in parentheses, $e_1$, is
basically the spread of the approximants for $\beta_c$ fixed at its MC
estimate.  It is the standard deviation of the results obtained from
all ``good'' IA's divided by the square root of $r_a$, i.e.,
$e_1=\sigma/\sqrt{r_a}$. The number in
brackets, $e_2$, is related to the uncertainty on the value of
$\beta_c$ and it is estimated by varying $\beta_c$ in the range
$[\beta_c-\Delta\beta_c,\beta_c+\Delta\beta_c]$.

\section{Universal amplitude ratios from the parametric representation} 
\label{univparrep}

In the following we report the expressions of the universal amplitude ratios 
in terms of the parametric representation (\ref{parrep}) of
the critical equation of state.

The singular part of the free energy per unit volume can be written as
\begin{equation}
{\cal F}_{\rm sing} = h_0 m_0 R^{2-\alpha} g(\theta),
\end{equation}
where $g(\theta)$ is the solution of the first-order differential
equation
\begin{equation}
(1-\theta^2) g'(\theta) + 2(2-\alpha)\theta g(\theta) = Y(\theta) h(\theta)
\label{pp1}
\end{equation}
that is regular at $\theta=1$.
The function $Y(\theta)$ has been defined in Eq.~(\ref{Yfunc}).
The longitudinal magnetic susceptibility can be written as
\begin{equation}
\chi_L^{-1} = {h_0\over m_0} R^\gamma g_2(\theta),\qquad\qquad
g_2(\theta) = {2\beta\delta \theta h(\theta) + (1-\theta^2) h'(\theta)\over Y(\theta)}.
\label{pp2}
\end{equation}
The function $g_2(\theta)$ must vanish at $\theta_0$ in order to reproduce the
predicted behavior at the coexistence curve $\chi_L \sim H^{-1/2}$, 
according to
\begin{equation}
g_2(\theta)\sim \theta_0-\theta\qquad\qquad{\rm for}\qquad \theta \to \theta_0.
\end{equation}
From Eq.~(\ref{pp2}) we
see that $g_2(\theta)$ satisfies this condition if $h(\theta)\sim (\theta_0-\theta)^2$
for $\theta\to \theta_0$.

From the equation of state one can derive universal amplitude ratios of
zero-momentum quantities. We consider
\begin{eqnarray}
&& U_0 \equiv A^+/A^-=
(\theta_0^2 - 1 )^{2-\alpha} {g(0)\over g(\theta_0)},\\
&& R_\chi \equiv {C^+ B^{\delta-1}\over B_c^\delta}=
(\theta_0^2-1)^{-\gamma} [m(\theta_0)]^{\delta-1} [m(1)]^{-\delta} h(1),\\
&& R_C \equiv {\alpha A^+ C^+\over B^2} =
- \alpha (1-\alpha)(2-\alpha) (\theta_0^2 - 1 )^{2\beta} [m(\theta_0)]^{-2} g(0),\\
&& R_4 \equiv  - {C_4^+ B^2\over (C^+)^3} =
\rho^2 \,[m(\theta_0)]^2 \left(\theta_0^2-1\right)^{-2\beta}.
\end{eqnarray}
Using Eqs.\ (\ref{fxmt}) and (\ref{Fzrel}) one can easily 
derive the expressions of the various coefficients that
characterize the asymptotic behavior of the scaling functions
$f(x)$ and $F(z)$, such as $c_f,f_i^0$ for $f(x)$ and
$F^\infty_i,r_{2j}$ for $F(z)$. 
Concerning the ratios involving amplitudes along the crossover line,
one finds
\begin{eqnarray}
&& P_m \equiv { T_p^\beta B\over B_c} = x_{\rm max}^\beta f(x_{\rm max})^{-1/\delta},\\ 
&& P_c \equiv - { T_p^{2\beta\delta} C^+ \over C_4} = F(z_{\rm max})^{-2},\\
&& R_p \equiv { C^+\over C_p} = F'(z_{\rm max}).
\end{eqnarray}
Here $x_{\rm max}$ and $z_{\rm max}$ are the values of the scaling 
variables $x$ and $z$ computed at $\theta_{\rm max}$,
where $\theta_{\rm max}$  is the solution of the equation 
\begin{equation}
\beta\delta F[z(\theta)] F''[z(\theta)] - \gamma F'[z(\theta)]^2 = 0.
\label{eq:C12}
\end{equation}


\begin{thebibliography}{99}
\bibitem{foot1}
The isotropic Heisenberg model is a simplified model 
and neglects several interactions that are present in real materials.
Among them, we should mention the cubic anisotropies due to the lattice 
structure and the dipolar interactions. Even if, 
in the renormalization-group language, these effects are relevant perturbations
of the Heisenberg fixed point 
\cite{Fisher-74,Aharony-76,AF-73,CPV-00},
the new critical exponents are so close to those of the 
Heisenberg universality class that the difference is experimentally
very difficult to observe, see, e.g., 
Refs.~\cite{BA-74,CPV-00,SKS-00,PV-review} 
and references therein.
\bibitem{foot2} For some dopings and some divalent cation A a first-order
transition has been observed. Moreover, in systems in which the transition 
appears to be of second order, mean-field critical exponents have been
measured. For instance, for La$_{1-x}$Sr$_x$MnO$_3$, a mean-field 
value for $\beta$ was observed in 
Refs.~\cite{LRKBMT-97,MSKMM-98,SSA-00},
while an estimate compatible with the Heisenberg value was found in
Refs. \cite{MSEHMT-96,HLHNLKNUMC-96,LBGGKSAM-97,GLGKSAM-98}.
For $x=1/3$ there also exists \cite{Ramos-etal-01} an estimate of the exponent
$\alpha$, $\alpha = - 0.14\pm 0.10$, in agreement with the 
Heisenberg value.
\bibitem{foot3}
In order to observe the correct exponents, it is essential to consider 
corrections to scaling in the analysis of the experimental data \cite{Kaul-85}.
All results reported in Table \ref{table_exponents_exp_2}, except 
those of Ref. \cite{SHAM-99}, have been obtained by assuming 
scaling corrections of the form $(1 + a |t|^{\Delta_1} + b 
|t|^{\Delta_2})$, with $\Delta_1 = 0.11$ and 
$\Delta_2 = 0.55$. Two observations are in order here. 
First, the results of the present paper provide a precise estimate 
of $\Delta_1$, $\Delta_1 = 0.1336(15)$. Second, the renormalization group
predicts also corrections 
of order $|t|^{2\Delta_1}$, $|t|^{3\Delta_1}$, etc. which are more 
relevant than the term $|t|^{\Delta_2}$ and should therefore be taken into 
account in the analysis of the data.
\bibitem{foot-doubleexchange}
Recently, a model with competing superexchange and 
double-exchange interactions has been studied 
\cite{TL-01}. A 
preliminary analysis for the paramagnetic-ferromagnetic
transition gives $\nu = 0.720(2)$ and $\gamma = 1.438(8)$.
While $\nu$ is reasonable agreement with the Heisenberg
value, $\gamma$ is significantly higher, so that the 
identification of this transition as a Heisenberg one is in doubt.
\bibitem{CPRM}
In the 
so-called critical-point renormalization method  
(see Ref.\ \cite{int-appr-ref} and references therein), given two
series $D(x)$ and $E(x)$ that are singular at the same point $x_0$,
$D(x)=\sum_i d_i x^i\sim (x_0-x)^{-\delta}$ and 
$E(x)=\sum_i e_i x^i\sim (x_0-x)^{-\epsilon}$, one constructs a new
series $F(x)=\sum_i (d_i/e_i)x^i$.  The function $F(x)$ is singular at
$x=1$ and for $x\to 1$ behaves as $F(x)\sim (1-x)^{-\phi}$, where
$\phi = 1+ \delta - \epsilon$.  Therefore, the difference
$\delta-\epsilon$ can be obtained by analyzing the expansion of $F(x)$
by means of biased approximants with a singularity at $x_c=1$.
\bibitem{foot-YL}
In the HT phase $t>0$, $Mt^{-\beta} = g(H t^{-\beta\delta})$
where $g(\tau)$ is analytic in the complex $\tau$ plane except 
possibly for two branch cuts on the imaginary axis, for 
$|{\rm Im}\ \tau| > \tau_0$ (Yang-Lee theorem). Thus,
the convergence radius is at most $|z| < |g(i\tau_0)|/k$, where
$k$ is the normalization of $z$.
\bibitem{foot-Rchi}
The amplitudes $B$, $B_c$, and $C^+$ are determined in several 
experiments---see, e.g., Ref. \cite{BK-97}---and thus an estimate of the 
ratio $R_\chi$ is possible. However, these amplitudes have large
systematic uncertainties, and we can only obtain a rough 
estimate, $0.6\lesssim R_\chi \lesssim 2$.
\bibitem{foot4}
We have evaluated $({u}_{\rm exp})_{\rm max}$ (the position of the 
maximum) from Fig. 4 of Ref.~\cite{Zhao-etal-99}:
$({u}_{\rm exp})_{\rm max} \approx 0.0005$. 
The variable $u$ defined in Sec. \ref{unratio} is related to 
${u}_{\rm exp}$ by ${u} = u_{\rm exp}/({u}_{\rm exp})_{\rm max}$.
Note that the $({u}_{\rm exp})_{\rm max}$ can also be derived from the 
knowledge of the nonuniversal constants $B$, $B_c$, and $C^+$, and 
using the results of Sec. \ref{unratio}. Indeed,
\[
({u}_{\rm exp})_{\rm max} = 
 \left({B_c\over B}\right)^{1/\beta} w_{\rm max} = 
 \left({C^+\over B_c}\right)^{1/\gamma} w_{\rm max} R_\chi^{-1/\gamma} = 
 \left({C^+\over B}\right)^{1/(\beta \delta)} w_{\rm max} 
  R_\chi^{-1/(\beta\delta)}.
\]
Numerically, using the estimates reported in Sec. \ref{unratio}, we have
$w_{\rm max} = 1.55(6)$, $w_{\rm max} R_\chi^{-1/\gamma} = 1.28(7)$,
and $w_{\rm max} R_\chi^{-1/(\beta\delta)} = 1.33(7)$.
\end{thebibliography}
\end{document}